\RequirePackage[2025-12-01]{latexrelease}
\documentclass[showpacs,showkeys,11pt,
preprint,preprintnumbers,nofootinbib,floatfix,
groupedaddress,superscriptaddress,amsmath,amssymb]{revtex4}

\usepackage{cancel}
\newcommand{\rev}[1]{#1}
\usepackage{amsfonts}
\usepackage{amsmath}
\usepackage{graphicx}
 \usepackage{array}
\graphicspath{{plots/}}
 \usepackage{subcaption} 
\usepackage{float}
 \usepackage{placeins}
\usepackage[export]{adjustbox}
\usepackage{verbatim}
\usepackage{epsfig}
\usepackage{url}
\usepackage{multirow}
\usepackage{hhline}
\usepackage{booktabs}
\usepackage{csquotes}

\usepackage[hidelinks]{hyperref}
\usepackage[nameinlink,noabbrev]{cleveref}
\crefname{section}{Sec.}{Secs.}
\Crefname{section}{Sec.}{Secs.}
\crefname{subsection}{Sec.}{Secs.}
\Crefname{subsection}{Sec.}{Secs.}
\crefname{subsubsection}{Sec.}{Secs.}
\Crefname{subsubsection}{Sec.}{Secs.}
\crefname{figure}{Fig.}{Figs.}
\Crefname{figure}{Figure}{Figures}
\crefname{table}{Table}{Tables}
\Crefname{table}{Table}{Tables}
\crefname{equation}{Eq.}{Eqs.}
\Crefname{equation}{Equation}{Equations}

\newcommand{\be}{\begin{equation}}
\newcommand{\ee}{\end{equation}}
\newcommand{\ba}{\begin{eqnarray}}
\newcommand{\ea}{\end{eqnarray}}

\begin{document}

\title{Probing Singlet Vector-Like Top Quarks in the Hadronic tZ Channel at the HL-LHC using Machine and Deep Learning Architectures }

\author{Haroon Sagheer}
\email{haroonsagheer663@gmail.com}
\affiliation{Riphah International University, Islamabad, Pakistan}

\author{M. Tayyab Javaid}
\email{Ch.tayyab119933@gmail.com}
\affiliation{Federal Urdu University of Arts, Science and Technology, Islamabad, Pakistan}

\author{Ijaz Ahmed}
\email{ijaz.ahmed@fuuast.edu.pk}
\affiliation{Federal Urdu University of Arts, Science and Technology, Islamabad, Pakistan}

\author{Ali Hassan}
\email{ali.hassan@phd.unict.it}
\email{ali.hassan@lns.infn.it}
\affiliation {Department of Physics and Astronomy "Ettore Majorana", University of Catania, Piazza Università, Italy}
\affiliation{INFN - Istituto Nazionale di Fisica Nucleare, S. Sofia, 95123, Catania, Italy}
\date{\today}

\author{Farzana Ahmad}
\email{farzana@konkuk.ac.kr}
\affiliation{SERI, \& College of Engineering, Konkuk University, Seoul 05029, South Korea}
\begin{abstract}
\rev{We study single production of a singlet vector-like top partner $T$ at the 14~TeV HL-LHC in the channel $pp\to Tj$, with $T\to tZ$, $t\to bjj$ and $Z\to\nu\bar\nu$, for $m_T$ = 1.8~-~3.1~TeV at the benchmark $g^*=0.4$, $R_L=0.5$. We simulate the signal and the available $t\bar t$, $tZj$, $ZZjj$, and $WZjj$ backgrounds at leading order with detector effects, and classify them with XGBoost and a graph neural network (GNN). The original pipelines, which used different input representations, reach test AUCs of 0.976~-~0.983 (XGBoost) and 0.983--0.990 (GNN); these are not a like-for-like comparison. A controlled retraining with identical inputs, process-stratified partitions, seven masses and three seeds finds XGBoost ahead in all 21 paired runs, but only by $2\times10^{-4}$--$2.3\times10^{-3}$ in AUC, so no general ranking of architectures follows. At $3000~\mathrm{fb}^{-1}$, the statistical Asimov significance reaches at most 3.4 (XGBoost) and 3.8 (GNN) and falls below 0.5 with a 20\% background-normalization uncertainty. An audit of the original workflow identified cross-mass validation training overlap, inconsistent yield normalization, and limited background statistics after stringent selections. Consequently, coupling-dependent contours are not reported. Because several relevant background processes are not simulated, the quoted significances are conditional on the specified background model and should not be interpreted as an HL-LHC exclusion or discovery reach.}
\end{abstract}
%We study single production of a singlet vector-like top partner \(T\) at the 14~TeV HL-LHC in the channel \(pp\to Tj\), with \(T\to tZ\), \(t\to bW\to bjj\), and \(Z\to\nu\bar\nu\). \rev{Because the historical XGBoost and graph-neural-network (GNN) pipelines used different input representations, their respective test-AUC ranges of 0.976--0.983 and 0.983--0.990 are reported as pipeline-specific results rather than as a controlled comparison of architectures. We therefore perform an additional comparison using identical physical information, common process-stratified training, validation, and test partitions, and three training seeds.} We evaluate statistical Asimov sensitivities and a variant containing a 20\% background-normalization uncertainty. \rev{Within the four-process background model of the original analysis, this uncertainty reduces the benchmark significance to below 0.5 throughout the mass scan, illustrating its strong dependence on the assumed background normalization.} \rev{The quantitative sensitivity analysis is restricted to the directly simulated benchmark \(g^*=0.4,\ R_L=0.5\); no \(g^*\)-dependent exclusion or discovery contours are reported.} \rev{The quoted significances are therefore conditional on the simulated \(t\bar t\), \(tZj\), \(ZZjj\), and \(WZjj\) backgrounds and do not constitute a complete experimental HL-LHC projection.}

%\end{abstract}

%\vspace{0.8cm}
%\noindent\textbf{Emails:}\\
%\texttt{haroonsagheer663@gmail.com}\\
%\texttt{ch.tayyab119933@gmail.com}\\
%\texttt{ijaz.ahmed@fuuast.edu.pk}\\
%\texttt{mjamil@konkuk.ac.kr}

%\vspace{0.5cm}
%\noindent\textbf{PACS:} 12.60.Fr, 14.65.Jk, 12.60.-%i, 14.80.Fd\\
%\noindent\textbf{Keywords:} Vector-like quarks; HL-LHC; XGBoost; GNN; Machine learning; Signal significance
\maketitle
%\end{titlepage}
%\setcounter{page}{1}
%%%%%%%%%%%%%%%%%%%%%%%%%%%%%%%%%%%%%%%%%%%%%%%%%%%%%%%%%%%%%%%%%%%
\section{Introduction}
\label{sec:intro}
%%%%%%%%%%%%%%%%%%%%%%%%%%%%%%%%%%%%%%%%%%%%%%%%%%%%%%%%%%
The Standard Model (SM) describes existing collider measurements with remarkable precision, yet it leaves open basic questions about electroweak naturalness, flavor, and the structure of physics at the TeV scale. A broad class of extensions predicts vector-like quarks (VLQs), heavy color-triplet fermions whose left- and right-handed components have the same gauge quantum numbers \cite{AguilarSaavedra2013,Buchkremer2013,Cacciapaglia2012}. Such states arise in composite Higgs models, warped extra dimensions, and little Higgs theories, where gauge-invariant mass terms and mixing with SM quarks can help address Higgs-sector radiative corrections \cite{PanicoWulzer2016,RandallSundrum1999,ArkaniHamed2002LH}.

For a singlet top partner \(T\), the dominant decay channels are \(T\to bW\), \(T\to tZ\), and \(T\to th\), with asymptotic branching ratios dictated by Goldstone equivalence at high mass \cite{AguilarSaavedra2013,Cacciapaglia2010Widths}. At the LHC, pair production is QCD-driven and approximately independent of electroweak mixing, but rapidly loses phase space at multi-TeV masses. Single production is model dependent and directly sensitive to the effective mixing coupling. Dedicated phenomenological studies have investigated the \(tZ\), \(Z\to\nu\bar\nu\) channel at the LHC and future hadron colliders \cite{LiChaoZhang2023,Han2023tZ}.

ATLAS and CMS searches for singly produced \(T\) quarks in \(tZ\) and \(tH\) final states already push singlet-\(T\) limits into the TeV range \cite{CMS2017SingleVLQZ,CMS2018SingleTtZ,CMS2022tZMET,ATLAS2023tZtH,CMS2024tHtZ,ATLAS2025SingleTCombination}. Particularly relevant to the present topology, CMS used \(137~\mathrm{fb}^{-1}\) of 13~TeV data in the jets-plus-missing-momentum final state and set observed narrow-width limits on \(\sigma\mathcal{B}(T\to tZ)\) ranging from 602 to 15~fb over \(m_T=0.6\)--1.8~TeV \cite{CMS2022tZMET}. ATLAS subsequently used an XGBoost-based boosted-top classifier in the same broad top-plus-missing-momentum signature and excluded \(m_T<1.8~\mathrm{TeV}\) for \(\kappa_T=0.5\) and \(\mathcal{B}(T\to tZ)=25\%\) \cite{ATLAS2024MonoTopXGB}. A later ATLAS combination of the \(Ht\) and \(Zt\) channels excluded masses below 2.1~TeV for the same singlet benchmark \cite{ATLAS2025SingleTCombination}. These experimental results use profile-likelihood inference and data-constrained backgrounds; their \(\kappa_T\) convention is not automatically identical to the \(g^*\) convention used here. The HL-LHC target luminosity of \(\sim3000~\mathrm{fb}^{-1}\) can extend the reach, but only if complex hadronic backgrounds are controlled \cite{HLLHC2015,Banerjee2024VLQReview}.

We therefore focus on the single-production topology \(pp\to Tj\) with \(T\to tZ\), \(t\to bW\to bjj\), and \(Z\to\nu\bar\nu\). The hadronic top decay provides a large branching fraction, while the invisible \(Z\) yields a missing-momentum handle. \rev{Following the baseline of Ref.~\cite{LiChaoZhang2023}, the available background model comprises \(t\bar t\), \(tZj\), \(ZZjj\), and \(WZjj\), including charge-conjugate modes. Other processes relevant to experimental top-plus-missing-momentum searches, including \(Z(\nu\bar\nu)+\)heavy-flavor jets, lost-lepton \(W+\)jets, single-top production, \(t\bar tZ\), \(t\bar tW\), and instrumental multijet backgrounds are absent from the available samples and therefore from the quoted yields. They are not assumed to be negligible; all reported yields and significances are conditional on the four-process background model.} Heavy-\(T\) kinematics induce correlated structures in jet momenta, angular separations, and event-level variables, motivating multivariate methods beyond a cut-based approach.

Simulation samples were generated with \textsc{MadGraph5\_aMC@NLO} v3.5.11 and processed with \textsc{Pythia8} and \textsc{Delphes} \cite{Alwall2014MG5,Sjostrand2015Pythia,Favereau2014Delphes}. \rev{After the hadronic preselection (\(N_j\ge3\), \(N_b\ge1\), \(N_\ell=0\)) and kinematic cuts, the historical analysis applied XGBoost \cite{Chen2016XGBoost} and a GNN based on jet nodes and relational edges \cite{Shlomi2021GNN,Zaheer2017DeepSets,Komiske2019EFN,Qu2020ParticleNet,Qu2022ParticleTransformer,Radovic2018ML}. Because the pipelines used different input representations, their performance metrics do not provide a controlled comparison of classifier architectures. A controlled study with identical physical information and common event partitions is presented in \cref{new:sec:analysis}.}

Compared with the cut-based study of Ref.~\cite{LiChaoZhang2023}, the original analysis adds XGBoost and GNN pipelines to the same broad 5FS detector-level topology and applies a common score threshold across the stored mass points. XGBoost itself is not a novel ingredient for this signature because ATLAS has already used it in a search for a boosted top quark plus missing transverse momentum \cite{ATLAS2024MonoTopXGB}. \rev{The original XGBoost and GNN results are therefore treated as outputs of two independent pipelines. Their AUC and sensitivity values are descriptive, and their numerical differences cannot be attributed solely to the classifier architectures or input representations. The controlled retraining described in \cref{new:sec:analysis} instead compares the classifiers using identical physical information.} \rev{The surviving outputs retain the AUC values and statistical sensitivities at the nominal benchmark but lack the normalization information required for a controlled \(g^*\)-dependent reinterpretation. The original rate contours are therefore not used to estimate the physical sensitivity or mass reach.}

\rev{The model is parameterized by \((m_T,g^*,R_L)\), but the quantitative sensitivity results are restricted to the nominal sample at \(g^*=0.4\) and \(R_L=0.5\). We do not quote \(g^*\)-dependent crossings of \(Z_A=2\) or \(Z_A=5\), because the recorded normalization does not permit a self-consistent coupling rescaling.} \rev{Although the original pipelines show strong test-sample discrimination, their different inputs preclude an architecture-level ranking. The controlled study compares only the two specified configurations supplied with identical physical information and does not imply a general ranking of the architectures.} \rev{The physics interpretation is therefore limited to the nominal benchmark and the simulated background model; the conditional significances are not presented as an experimental reach over the full parameter space.}

The paper is organized as follows. \Cref{sec:model} summarizes the singlet-\(T\) model and parameter definitions. \Cref{sec:simulation} details event generation, detector simulation, and selection cuts. \Cref{sec:analysis} presents the multivariate analysis, including model architectures and feature sets. \rev{\Cref{sec:results} presents the classifier performance and benchmark sensitivities.} \Cref{sec:conclusion} concludes with a summary and outlook.

%%%%%%%%%%%%%%%%%%%%%%%%%%%%%%%%%%%%%%%%%%%%%%%%%%%%%%%%%%%%%%%%%
\section{The Singlet Vector-like Top-quark Model}
\label{sec:model}
%%%%%%%%%%%%%%%%%%%%%%%%%%%%%%%%%%%%%%%%%%%%%%%%%%%%%%%%%%%%%%%%%
We summarize the simplified singlet-\(T\) framework used in the analysis and fix our notation. The broken-phase interaction parameterization follows Refs.~\cite{Buchkremer2013,AguilarSaavedra2013,LiChaoZhang2023}.

\subsection{Field Content, Couplings, and Parameterization}
\label{subsec:model_couplings}
We consider a vector-like top partner \(T\) with electric charge \(+2/3\), transforming as an \(SU(2)_L\) singlet. The interactions relevant for single production and decay are controlled by an overall coupling strength \(g^*\) and a nonnegative mixing parameter \(R_L\). To distinguish amplitude factors from coupling-squared weights, we define
\begin{equation}
 \zeta_1 \equiv \frac{R_L}{1+R_L}, \qquad
 \zeta_3 \equiv \frac{1}{1+R_L}, \qquad
 \zeta_1+\zeta_3=1.
 \label{eq:zeta}
\end{equation}
The interaction amplitudes contain \(\sqrt{\zeta_i}\). Thus \(R_L=0\) corresponds to pure third-generation coupling, whereas \(R_L\to\infty\) gives pure first-generation coupling. Neglecting the light-quark masses, the effective interaction Lagrangian after electroweak symmetry breaking is
\begin{align}
\mathcal{L}_T 
&= \frac{g g^*}{2}\Bigg\{
\sqrt{\zeta_1}\left[
\frac{1}{\sqrt{2}}\bar T_L\gamma^\mu W^+_\mu d_L
+\frac{1}{2c_W}\bar T_L\gamma^\mu Z_\mu u_L
-\frac{m_T}{2m_W}\bar T_R h\,u_L
\right]\nonumber\\
&\qquad\quad+
\sqrt{\zeta_3}\left[
\frac{1}{\sqrt{2}}\bar T_L\gamma^\mu W^+_\mu b_L
+\frac{1}{2c_W}\bar T_L\gamma^\mu Z_\mu t_L
-\frac{m_T}{2m_W}\bar T_R h\,t_L
-\frac{m_t}{2m_W}\bar T_L h\,t_R
\right]\Bigg\}+\mathrm{h.c.}
\label{eq:LT}
\end{align}
Here \(g\) is the \(SU(2)_L\) gauge coupling, \(c_W\equiv\cos\theta_W\), and \(h\) is the physical Higgs field. \Cref{eq:LT} is a broken-phase effective interaction, not a gauge-invariant Lagrangian written in terms of the Higgs doublet. The mass factors in the scalar interactions are required by dimensional consistency; the \(m_t\) term supplies the opposite chirality for the third generation. The parameters \(g^*\) and \(R_L\) respectively control the overall interaction strength and the relative first- versus third-generation weights; sizable light-generation mixing has been considered in model-independent VLQ searches \cite{Atre2011VLQ}.

\subsection{Partial Widths, Branching Ratios, and Limiting Behavior}
\label{subsec:widths}
The dominant third-generation decay modes of a singlet \(T\) are
\begin{equation}
T \to bW, \qquad T \to tZ, \qquad T \to tH,
\label{eq:decay_modes}
\end{equation}
with \(T\to dW,uZ,uh\) opening when \(R_L\ne0\). The numerical model implementation should be used for finite-mass widths. It is important not to mix coupling conventions here. From \cref{eq:LT}, the heavy-mass vertex coefficients (neglecting \(m_t\)) are
\begin{equation}
C_{Wi}=\sqrt{\zeta_i}\frac{g g^*}{2\sqrt{2}},\qquad
C_{Zi}=\sqrt{\zeta_i}\frac{g g^*}{4c_W},\qquad
y_{hi}=\sqrt{\zeta_i}\frac{g g^*m_T}{4m_W}.
\label{eq:vertex_coefficients}
\end{equation}
Consequently, in the limit \(m_T\gg m_t,m_W,m_Z,m_h\), these written vertices give \cite{Buchkremer2013,Cacciapaglia2010Widths}
\begin{align}
\Gamma(T\to W d_i) &\simeq \zeta_i\frac{(g g^*)^2}{256\pi}\frac{m_T^3}{m_W^2},\\
\Gamma(T\to Z u_i) &\simeq \zeta_i\frac{(g g^*)^2}{512\pi}\frac{m_T^3}{m_Z^2c_W^2},\\
\Gamma(T\to h u_i) &\simeq \zeta_i\frac{(g g^*)^2}{512\pi}\frac{m_T^3}{m_W^2},
\label{eq:widths_approx}
\end{align}
where \(i=1,3\) and \((\zeta_1,\zeta_3)\) are given by \cref{eq:zeta}. Since \(m_Zc_W=m_W\) at tree level, the asymptotic ratio for each generation is
\begin{equation}
\Gamma(Wd_i):\Gamma(Zu_i):\Gamma(hu_i)=2:1:1.
\label{eq:width_ratio}
\end{equation}
The factors \(1/(64\pi)\) and \(1/(128\pi)\) quoted in Eqs.~(2)--(4) of Ref.~\cite{LiChaoZhang2023} are four times larger and do not follow from its displayed Lagrangian, which has the same overall vertex normalization as \cref{eq:LT}. They would correspond to vertices twice as large, or equivalently to a redefined coupling. Because the UFO model and parameter card used for the available samples are unavailable, the absolute mapping between their generator parameter and the \(g^*\) written here cannot be established. The ratio \(2:1:1\) and the branching-ratio limit below are unaffected by this common normalization ambiguity.

Summing the leading terms over both generations gives
\begin{equation}
\frac{\Gamma_T}{m_T}\simeq
\frac{(g g^*)^2}{128\pi}\frac{m_T^2}{m_W^2},
\label{eq:width_fraction}
\end{equation}
provided that no nonstandard decay modes are open. Taking \(g\simeq0.65\), \cref{eq:width_fraction} gives approximately 0.10 at \((m_T,g^*)=(2.0~\mathrm{TeV},0.4)\) and 0.39 at \((3.1~\mathrm{TeV},0.5)\). Thus a narrow-width rate rescaling is already marginal at the nominal benchmark and fails in the upper-right part of the stated scan. A regenerated analysis must use finite-width samples (including interference where relevant) or restrict the scan using an explicit \(\Gamma_T/m_T\) criterion \cite{Deandrea2021Width}.

\begin{figure}[!htbp]
  \centering
  \IfFileExists{f1.png}{%
    \includegraphics[width=0.78\linewidth]{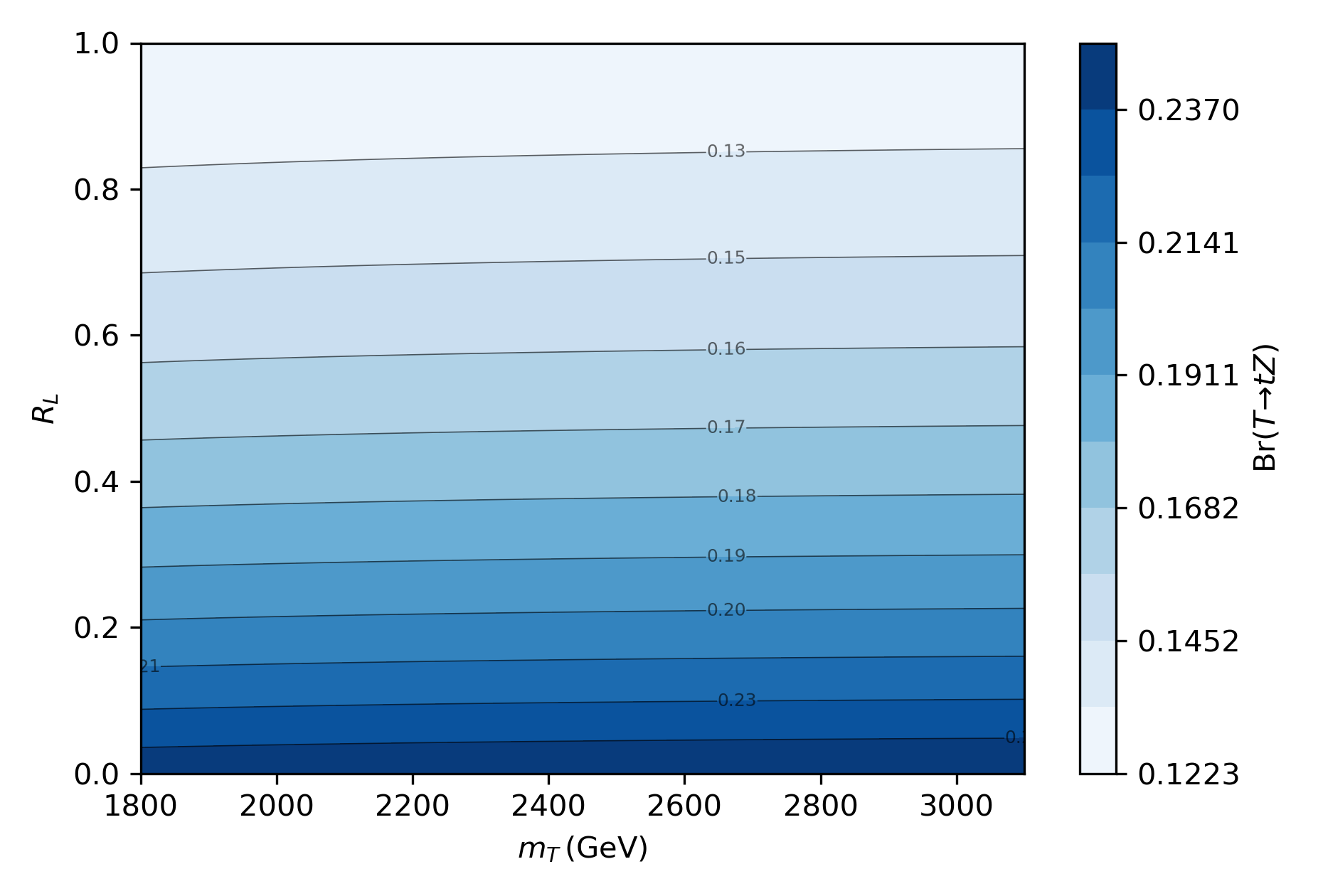}%
  }{%
    \fbox{\rule{0pt}{4cm}\rule{\dimexpr0.78\linewidth-2\fboxsep-2\fboxrule\relax}{0pt}}%
  }
\caption{Branching ratio $\mathrm{Br}(T\to tZ)$ in the $(R_L,m_T)$ plane.}
  \label{fig:br_tz}
\end{figure}

The branching ratio for the signal channel is
\begin{equation}
\mathrm{Br}(T\to tZ)=\frac{\Gamma(T\to Zt)}{\Gamma_{\rm tot}}.
\label{eq:br_definition}
\end{equation}
In the heavy-mass limit the overall factor \((g g^*)^2\) cancels, and \cref{eq:zeta,eq:widths_approx} imply
\begin{equation}
\mathrm{Br}(T\to tZ)\longrightarrow\frac{\zeta_3}{4}
=\frac{1}{4(1+R_L)}.
\label{eq:br_limit}
\end{equation}
Finite-mass phase space produces the weak residual \(m_T\) dependence visible in \cref{fig:br_tz}.

For \(R_L=0\), couplings are purely to the third generation and \(\mathrm{Br}(T\to tZ)\) is maximal within the singlet framework. As \(R_L\) increases, the coupling is shared with the first generation, reducing \(\mathrm{Br}(T\to tZ)\) while opening light-quark modes. The contour map therefore provides a direct link between the model parameter \(R_L\) and the observable \(tZ\) signal fraction used in the analysis.

\subsection{Benchmark Choices}
\label{subsec:benchmarks}
The recorded scan contains \(R_L=0\) and \(R_L=0.5\) labels and \(g^*\in[0.1,0.5]\). The mass points are \(m_T=1.8\), 2.0, 2.2, 2.5, 2.7, 2.9, and 3.1~TeV.

We adopt the five-flavor scheme (5FS), in which the \(b\) quark is treated as a parton in the proton. This is standard for single production of heavy top partners and enhances the relevance of the \(Tj\) topology at large masses. Details of event generation, PDF choice, and matching are given in \cref{sec:simulation}.

At fixed \(R_L\), single production scales approximately quadratically with \(g^*\) in the narrow-width regime:
\begin{equation}
\sigma(pp\to Tj)\,\mathrm{Br}(T\to tZ)
\propto (g^*)^2\,F(m_T,R_L).
\label{eq:rate_scaling}
\end{equation}
This rescaling also assumes that the acceptance and classifier response are insensitive to \(g^*\), which is justified only while the narrow-width approximation remains valid. Crucially, changing \(R_L\) requires the full production factor \(F(m_T,R_L)\), including the different light- and heavy-quark parton luminosities; rescaling only \(\mathrm{Br}(T\to tZ)\) is not valid.
%%%%%%%%%%%%%%%%%%%%%%%%%%%%%%%%%%%%%%%%%%%%%%%%
\section{Event Selection and Kinematic Distribution}
\label{sec:simulation}
%%%%%%%%%%%%%%%%%%%%%%%%%%%%%%%%%%%%%%%%%%%%%%%%
Signal and background samples were generated at LO with \textsc{MadGraph5}\allowbreak\_\allowbreak\textsc{aMC@NLO} v3.5.11 \cite{Alwall2014MG5}. The signal process is
\begin{equation}
pp\to Tj,\qquad T\to tZ,\qquad t\to bW\to bjj,\qquad Z\to\nu\bar{\nu},
\label{eq:signal_process}
\end{equation}
including the charge-conjugate mode \(pp\to \bar{T}j\). The available background samples are \(t\bar t\), \(tZj\), \(ZZjj\), and \(WZjj\), including charge-conjugate channels. Signal samples were generated at the nominal point \(g^*=0.4\), \(R_L=0.5\). Rescaling in \(g^*\) at this fixed \(R_L\) can use the quadratic relation in \cref{sec:model} within the narrow-width approximation. The original \(R_L=0\) reinterpretation cannot be validated because neither an \(R_L\)-dependent production cross-section grid nor the original reweighting code survives. We therefore do not interpret the two \(R_L\) contour plots as physical limits. The analysis uses the five-flavor scheme (5FS), in which the \(b\) quark is a proton parton.

Parton-level generation cuts are imposed to stabilize the matrix-element integration and reflect basic detector acceptance:
\begin{align}
&p_T(j),\,p_T(b),\,p_T(\ell) > 25~\mathrm{GeV},\nonumber\\
&|\eta(j)|<5.0,\quad |\eta(b)|<5.0,\quad |\eta(\ell)|<2.5,\nonumber\\
&\Delta R(x,y) > 0.4\quad (x,y\in\{j,b,\ell\}\ \text{where applicable}).
\label{eq:generation_cuts}
\end{align}
The LHAPDF6 identifier 331100 corresponds to \texttt{NNPDF40\_nnlo\_as\_01180} \cite{Buckley2015LHAPDF,NNPDF40}. This NNLO PDF was used with LO matrix elements in the original event generation; a regenerated analysis should use a perturbatively matched PDF choice and propagate PDF and scale uncertainties. The manuscript notes record
\(m_t=172.52\)~GeV, \(m_h=125.11\)~GeV, \(m_Z=91.1876\)~GeV, \(m_W=80.379\)~GeV, \(\alpha_s(m_Z)=0.118\), \(1/\alpha(m_Z)=132.185\), and \(G_F=1.1663787\times10^{-5}~\mathrm{GeV}^{-2}\) \cite{PDG2024}. The original parameter card is unavailable, so the electroweak input scheme and the values actually used by the generator cannot be independently verified.

Parton showering and hadronization were performed with \textsc{Pythia8} \cite{Sjostrand2015Pythia}. Detector effects were simulated with \textsc{Delphes} \cite{Favereau2014Delphes}; jets were reconstructed with the anti-\(k_T\) algorithm and radius parameter \(R=0.4\) \cite{Cacciari2008AntiKt,Cacciari2012FastJet}. The stated CMS-style Delphes card is not among the available files, preventing verification of the jet, \(b\)-tag, lepton-veto, and missing-momentum efficiencies. No additional multijet merging was applied to the signal; the recoil jet was generated at matrix-element level in \(pp\to Tj\).

After detector simulation, events must satisfy a hadronic preselection:
\begin{equation}
N_j\ge3,\qquad N_b\ge1,\qquad N_\ell=0.
\label{eq:preselection}
\end{equation}
Additional recorded cuts are then applied to suppress backgrounds while retaining signal efficiency in the boosted regime. The baseline selections listed in the original analysis are:
\begin{align}
&p_T(b_1) > 60~\mathrm{GeV},\qquad p_T(j_1) > 120~\mathrm{GeV},\nonumber\\
&\Delta R(j_1,j_2) < 3.0,\qquad \Delta R(j_1,j_3) < 3.5,\nonumber\\
&E_T^{\rm miss} > 160~\mathrm{GeV},\nonumber\\
&40~\mathrm{GeV} < m_{jj} < 140~\mathrm{GeV},\qquad \chi^2_{t} < 60,\nonumber\\
&80~\mathrm{GeV} < m_{bjj} < 320~\mathrm{GeV}.
\label{eq:selection_cuts}
\end{align}
These requirements target the reconstructed hadronic top and \(W\) candidates while maintaining sensitivity to heavy-\(T\) kinematics. The full list of observables used in the multivariate step is described in \cref{sec:analysis}.

\Cref{tab:cutflow_bench} summarizes the surviving cutflow for \(m_T=1800\) and 2000~GeV and the four available background samples. \rev{Earlier manuscript notes assigned \(K\) factors of 1.27, 1.14, and 1.10 to \(ZZjj\), \(WZjj\), and \(tZj\), respectively. Direct inspection of the recovered event weights instead gives \(K=1.15\) for \(ZZjj\); the \(WZjj\) and \(tZj\) factors remain 1.14 and 1.10, respectively. The \(t\bar t\) sample uses a factor of 1.616 associated with the NNLO QCD normalization \cite{Czakon2013ttbar}, and the signal factor is 0.95. These factors are retained solely to reproduce the original sample normalization and are not interpreted as universal higher-order corrections \cite{Fuks2017VLQNLO,Cacciapaglia2019Kfactor}.} Because the original run cards and complete upstream weight bookkeeping are absent, the generator-level normalization cannot be independently regenerated from the available files. All stored sensitivity plots assume \(3000~\mathrm{fb}^{-1}\). \rev{The absolute yields and significances thus refer only to the sum of the four available background samples. Processes absent from the dataset are excluded from this sum but are not assumed to be negligible.}

\begin{table}[H]
\centering
\caption{Recorded cutflow for the signal benchmarks (T1800, T2000) and four simulated backgrounds at 14~TeV. Entries are recorded as effective cross sections in fb; their normalization is not independently reproducible from the surviving files.}
\label{tab:cutflow_bench}
\resizebox{0.8\textwidth}{!}{%
\begin{tabular}{|c|c|c|c|c|c|}
\hline
Cuts & Signal (fb) & \multicolumn{4}{c}{Backgrounds (fb)} \\
\cline{3-6}
 & T1800(T2000) & $tt$ & $tZj$ & $ZZjj$ & $WZjj$ \\
\hline
$\sigma$ (Before cut) & 9.832(6.679) & 340362.082 & 44.022 & 216.660 & 548.331 \\
Trigger & 5.089(3.246) & 140370.768 & 23.131 & 50.246 & 69.797 \\
$p_T[b_1] > 60\,\mathrm{GeV}$ & 4.489(2.846) & 97818.020 & 14.093 & 27.919 & 39.759 \\
$p_T[j_1] > 120\,\mathrm{GeV}$ & 4.438(2.818) & 39383.296 & 6.094 & 13.426 & 23.778 \\
$\Delta R[j_1,j_2] < 3.0$ & 2.741(1.707) & 22951.977 & 2.243 & 8.074 & 13.971 \\
$\Delta R[j_1,j_3] < 3.5$ & 1.598(0.982) & 20713.075 & 1.818 & 7.287 & 12.373 \\
$E_T^{\mathrm{miss}} > 160\,\mathrm{GeV}$ & 1.590(0.978) & 3360.054 & 0.750 & 3.135 & 5.769 \\
N(light jets) $\ge 2$ & 1.546(0.951) & 3055.771 & 0.675 & 2.670 & 5.497 \\
$40 < m_{jj} < 140\,\mathrm{GeV}$ & 0.890(0.542) & 2415.890 & 0.401 & 1.582 & 3.249 \\
$\chi^2_{\mathrm{top}} < 60$ & 0.639(0.371) & 2332.842 & 0.386 & 1.360 & 2.686 \\
$80 < m_{bjj} < 320\,\mathrm{GeV}$ & 0.585(0.336) & 2282.468 & 0.379 & 1.288 & 2.526 \\
\hline
\end{tabular}%
}
\end{table}

\begin{figure}[h]
  \centering
  \IfFileExists{feynman.pdf}{%
    \includegraphics[width=0.45\linewidth]{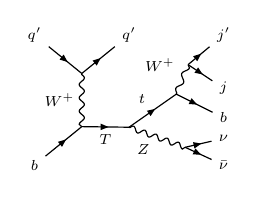}%
  }{%
    \fbox{\rule{0pt}{5cm}\rule{\dimexpr0.95\linewidth-2\fboxsep-2\fboxrule\relax}{0pt}}%
  }
\caption{Representative LO \(t\)-channel diagram for single production \(pp\to Tj\) in the 5FS with \(T\to tZ\), \(t\to bW\to bjj\), and \(Z\to\nu\bar{\nu}\).}
  \label{fig:2}
\end{figure}

%%%%%%%%%%%%%%%%%%%%%%%%%%%%%%%%%%%%%%%%%%%%%%%%%%%%%%%%%%%%%%%%%%%%%%%%
\section{Advanced Machine-learning Analysis for Classification}
\label{sec:analysis}
%%%%%%%%%%%%%%%%%%%%%%%%%%%%%%%%%%%%%%%%%%%%%%%%%%%%%%%%%%%%%%%%%%%%%%%%
The hadronic \(tZ\) topology with missing transverse momentum exhibits correlated kinematic and angular structure that can benefit from multivariate classification. \rev{The historical study used two independently configured pipelines: XGBoost with engineered tabular features and a GNN that represented each event as a jet graph. Because the feature representations differ, the observed performance differences cannot be attributed to the architectures alone. The controlled common-input comparison introduced in \cref{new:sec:analysis} is documented separately.} Machine-learning methods are now standard tools in collider analyses \cite{Baldi2014DeepLearning,Radovic2018ML,Shlomi2021GNN,Komiske2019EFN,Qu2020ParticleNet,Qu2022ParticleTransformer}.

For XGBoost we use the standard regularized boosting objective \cite{Chen2016XGBoost},
\begin{equation}
\mathcal{L} = \sum_i \ell\!\left(y_i,\hat{y}_i\right) + \sum_{k=1}^{K}\Omega(f_k),
\qquad \Omega(f_k)=\gamma T_k + \frac{\lambda}{2}\|w_k\|^2,
\label{eq:xgb_objective}
\end{equation}
where \(\ell\) is the logistic loss, \(f_k\) denotes a tree with \(T_k\) leaves and weights \(w_k\), and \(\gamma,\lambda\) control model complexity. The model predicts a signal score \(s(\mathbf{x})\in[0,1]\) from the feature vector \(\mathbf{x}\) and provides interpretable feature-importance measures. For the GNN, we construct an event graph in which jets are nodes and pairwise relations are encoded as edges, following permutation-invariant set and graph formulations \cite{Zaheer2017DeepSets,Shlomi2021GNN}. A generic message-passing layer updates node embeddings as
\begin{equation}
h_v^{(l+1)} = \phi\!\left(h_v^{(l)}, \sum_{u\in \mathcal{N}(v)} \psi\!\left(h_v^{(l)}, h_u^{(l)}, e_{vu}\right)\right),
\label{eq:message_passing}
\end{equation}
followed by a graph-level readout \(h_G=\mathrm{READOUT}(\{h_v\})\) and a classification head. \rev{This equation represents only the generic message-passing structure of the historical GNN. The exact edge construction, aggregation operator, preprocessing procedure, and software versions cannot be recovered from the original manuscript and figures. Consequently, the controlled GNN used for the referee-requested comparison is specified independently in \cref{new:sec:analysis} and is not presented as a reproduction of the unavailable historical model.}

\Cref{tab:features} summarizes the recorded inputs. The hadronic top candidate combines the leading \(b\)-tagged jet with the two highest-\(p_T\) light jets. We define \(H_T\) as the scalar sum of jet transverse momenta and \(S_T=H_T+E_T^{\rm miss}\). For the \(bjj\) system, a dimensionally unambiguous cluster transverse mass is
\begin{align}
m_T(bjj,E_T^{\rm miss})&=\sqrt{\left(E_T^{bjj}+E_T^{\rm miss}\right)^2
-\left|\vec p_T^{\,bjj}+\vec p_T^{\,\rm miss}\right|^2},\\
E_T^{bjj}&=\sqrt{m_{bjj}^2+\left|\vec p_T^{\,bjj}\right|^2}.
\label{eq:cluster_mt}
\end{align}
We also use \(\Delta\phi(\vec p_T^{\,bjj},\vec p_T^{\,\rm miss})\) and the balance variable \(H_T/(H_T+E_T^{\rm miss})\). The exact definition and resolution parameters of the top-\(\chi^2\) selection are absent from the available records; this definition must be restored before the cutflow is reproducible.

\begin{table}[!htbp]
\centering
\small
\caption{{Recorded feature sets for the historical independently configured XGBoost and GNN pipelines. Node features use the leading jets and the $b$ jet, edge features encode pairwise separations, and global features summarize event-level energy flow. The nonidentical historical inputs preclude an architecture-only interpretation of their recorded performance; the controlled same-information comparison is given in \cref{new:sec:analysis}.}}
\label{tab:features}
\resizebox{\textwidth}{!}{
\begin{tabular}{|c|c|c|}
\hline
\textbf{Category} & \textbf{XGBoost features} & \textbf{GNN features} \\
\hline
Global/event-level
& $E_T^{\rm miss}$, $S_T$, $H_T/(H_T+E_T^{\rm miss})$, $m_T(bjj,E_T^{\rm miss})$, $\Delta\phi(\vec{p}_T^{\,bjj}, \vec{p}_T^{\,\rm miss})$
& $E_T^{\rm miss}$, $H_T$, $S_T$, $H_T/(H_T+E_T^{\rm miss})$, $m_T(bjj,E_T^{\rm miss})$, $\Delta\phi(\vec{p}_T^{\,bjj}, \vec{p}_T^{\,\rm miss})$ \\
Reconstruction/topology
& $m_{bjj}$, $p_T(j_1)$
& --- \\
Angular structure
& $\Delta R(j_1,j_2)$, $\Delta R(j_1,j_3)$, $\Delta R(b,j_1)$
& $\Delta R(j_1,j_2)$, $\Delta R(j_1,j_3)$, $\Delta R(j_2,j_3)$, $\Delta R(b,j_1)$, $\Delta R(b,j_2)$ \\
Node features
& --- 
& $(p_T,\eta,\phi,m)$ for $j_1,j_2,j_3,b_1$ \\
\hline
\end{tabular}
}
\end{table}
\rev{The manuscript states that the historical pipelines used training, validation, and test subsets, with training weights rebalanced to equalize the total weighted signal and background contributions. The surviving outputs are insufficient to verify the original partitions or post-training normalization. The Kolmogorov--Smirnov (KS) tests reported in \cref{tab:ks_test} therefore provide only limited diagnostics of the historical models \cite{Massey1951KS}. The controlled comparison in \cref{new:sec:analysis} instead uses explicit process-stratified partitions, common preprocessing, multiple training seeds, stored score distributions, and validation-based early stopping.}

\rev{The recorded hyperparameters of the final historical models are summarized in \cref{tab:hyperparams}. Because the software versions and unspecified default settings were not preserved, these parameters alone are insufficient to reproduce the classifiers. The controlled comparison uses the fully specified implementations and software versions documented in \cref{new:sec:analysis}.}

\begin{table}[!htbp]
\centering
\small
\caption{Key hyperparameters used for the XGBoost and GNN models.}
\label{tab:hyperparams}
\begin{tabular}{|c|c|c|}
\hline
\textbf{Model} & \textbf{Hyperparameter} & \textbf{Value} \\
\hline
XGBoost & $n_{\rm estimators}$, $\max\,\mathrm{depth}$, $\eta$ & 350, 3, 0.05 \\
XGBoost & $\mathrm{subsample}$, $\mathrm{colsample\_bytree}$ & 0.7, 0.7 \\
XGBoost & $\gamma$, $\min\,\mathrm{child\_weight}$ & 0.2, 2.0 \\
XGBoost & $\lambda$, $\alpha$ & 1.0, 0.0 \\
GNN & Layers, hidden size, activation & 2, 64, ReLU \\
GNN & Dropout, optimizer & 0.2, Adam \\
GNN & Learning rate, weight decay & $10^{-3}$, $10^{-4}$ \\
GNN & Epochs, batch size & 50, 512 \\
\hline
\end{tabular}
\end{table}

Classifier performance is quantified using ROC curves and AUC. The output score \(s(\mathbf{x})\) is used as a discriminant. For a chosen threshold \(s>s_0\) we compute the weighted signal and background yields by
\begin{equation}
S(s_0)=\sum_{i\in \text{sig}} w_i\,\Theta(s_i-s_0),\qquad
B(s_0)=\sum_{i\in \text{bkg}} w_i\,\Theta(s_i-s_0),
\label{eq:weighted_yields}
\end{equation}
where $w_i$ are the event weights and $\Theta$ is the Heaviside step function. We report the Asimov significance \cite{Cowan2011Asimov,Cranmer2015PracticalStats},
\begin{equation}
Z_A = \sqrt{2\left[(S+B)\ln\left(1+\frac{S}{B}\right)-S\right]},
\label{eq:asimov}
\end{equation}
the simple estimator
\begin{equation}
Z_{S/\sqrt{S+B}} = \frac{S}{\sqrt{S+B}},
\label{eq:simple_significance}
\end{equation}
and an Asimov variant including a 20\% background systematic uncertainty $\sigma_B=0.2B$,
\begin{equation}
Z_{A,\,\sigma_B} = \sqrt{2\left[(S+B)\ln\left(\frac{(S+B)(B+\sigma_B^2)}{B^2+(S+B)\sigma_B^2}\right) - \frac{B^2}{\sigma_B^2}\ln\left(1+\frac{\sigma_B^2 S}{B(B+\sigma_B^2)}\right)\right]}.
\label{eq:asimov_systematic}
\end{equation}
\rev{The surviving records specify a threshold of \(s_0=0.99\) for the historical yield and significance calculations. Because the two classifiers have different score calibrations, this common numerical threshold does not define equivalent operating points and is not used for cross-pipeline comparison. The controlled study instead evaluates fixed signal-efficiency points and validation-selected thresholds using the recovered event-level scores; see \cref{new:tab:fixed,new:tab:wp}.} The \(Z_A=2\) and \(Z_A=5\) crossings shown later are statistical sensitivity proxies. A publishable exclusion requires a specified likelihood, nuisance parameters, correlations, and a confidence-level construction such as \(CL_s\) \cite{Read2002CLs}.
%%%%%%%%%%%%%%%%%%%%%%%%%%%%%%%%%%%%%%%%%%%%%%%%%%%%%%%%%%%%%%%%%%%%%%%%
\section{Results and Discussion}
\label{sec:results}
%%%%%%%%%%%%%%%%%%%%%%%%%%%%%%%%%%%%%%%%%%%%%%%%%%%%%%%%%%%%%%%%%%%%%%%%
\Cref{tab:cutflow_bench} (\cref{sec:simulation}) shows the cutflow for two representative benchmarks, \(m_T=1800\) and \(2000\)~GeV, together with the dominant backgrounds. The pattern follows the expected physics of the hadronic \(tZ\) topology. The \(E_T^{\rm miss}\) requirement targets the invisible \(Z\to\nu\bar{\nu}\) decay, while the hard recoil and collimated jets from a boosted top motivate the \(p_T(j_1)\) and \(\Delta R\) selections. The \(m_{jj}\) window and \(\chi^2_t\) enforce consistency with a hadronic \(W\to jj\) and \(t\to bjj\) reconstruction. Quantitatively (using the summed background columns), the trigger already removes \(\sim59\%\) of the background while keeping \(\sim49\)--\(52\%\) of the signal. The \(p_T(j_1)\) cut yields a further \(\sim60\%\) background reduction with \(\sim99\%\) signal retention, and the \(E_T^{\rm miss}\) cut is the single most powerful step, suppressing backgrounds by \(\sim84\%\) while retaining \(\gtrsim99.5\%\) of the signal. The two \(\Delta R\) cuts reduce the background by \(\sim42\%\) and \(\sim10\%\) per step with \(\sim58\)--\(62\%\) signal retention, and the \(m_{jj}\) window removes another \(\sim21\%\) of background at \(\sim57\%\) signal efficiency. The final top-consistency cuts (\(\chi^2_t\) and the \(m_{bjj}\) window) are milder (\(\sim3\%\) and \(\sim2\%\) background reduction) but improve the kinematic fidelity of the reconstructed top.\footnote{All quoted reductions and efficiencies are step-wise (relative to the preceding cut) and are consistent between the \(m_T=1800\) and \(2000\)~GeV benchmarks in \Cref{tab:cutflow_bench}.}

\begin{figure}[!htbp]
  \centering
  \IfFileExists{f3.png}{%
    \includegraphics[width=0.95\linewidth]{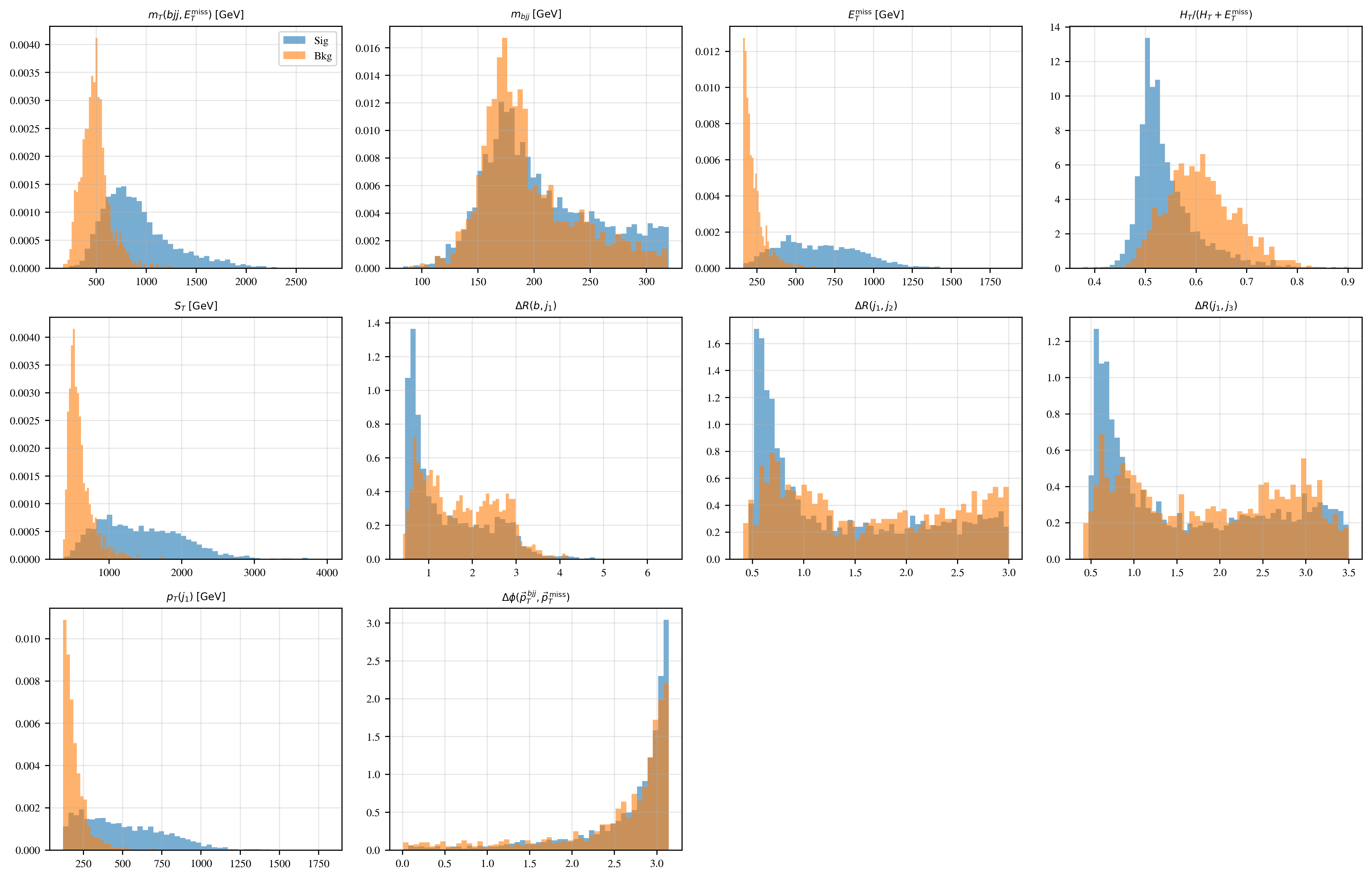}%
  }{%
    \fbox{\rule{0pt}{5cm}\rule{\dimexpr0.95\linewidth-2\fboxsep-2\fboxrule\relax}{0pt}}%
  }
\caption{Input variable distributions for XGBoost at the benchmark mass \(m_T=2000~\mathrm{GeV}\).}
  \label{fig:inputs_xgb}
\end{figure}

\begin{figure}[!htbp]
  \centering
  \IfFileExists{f4.png}{%
    \includegraphics[width=0.95\linewidth]{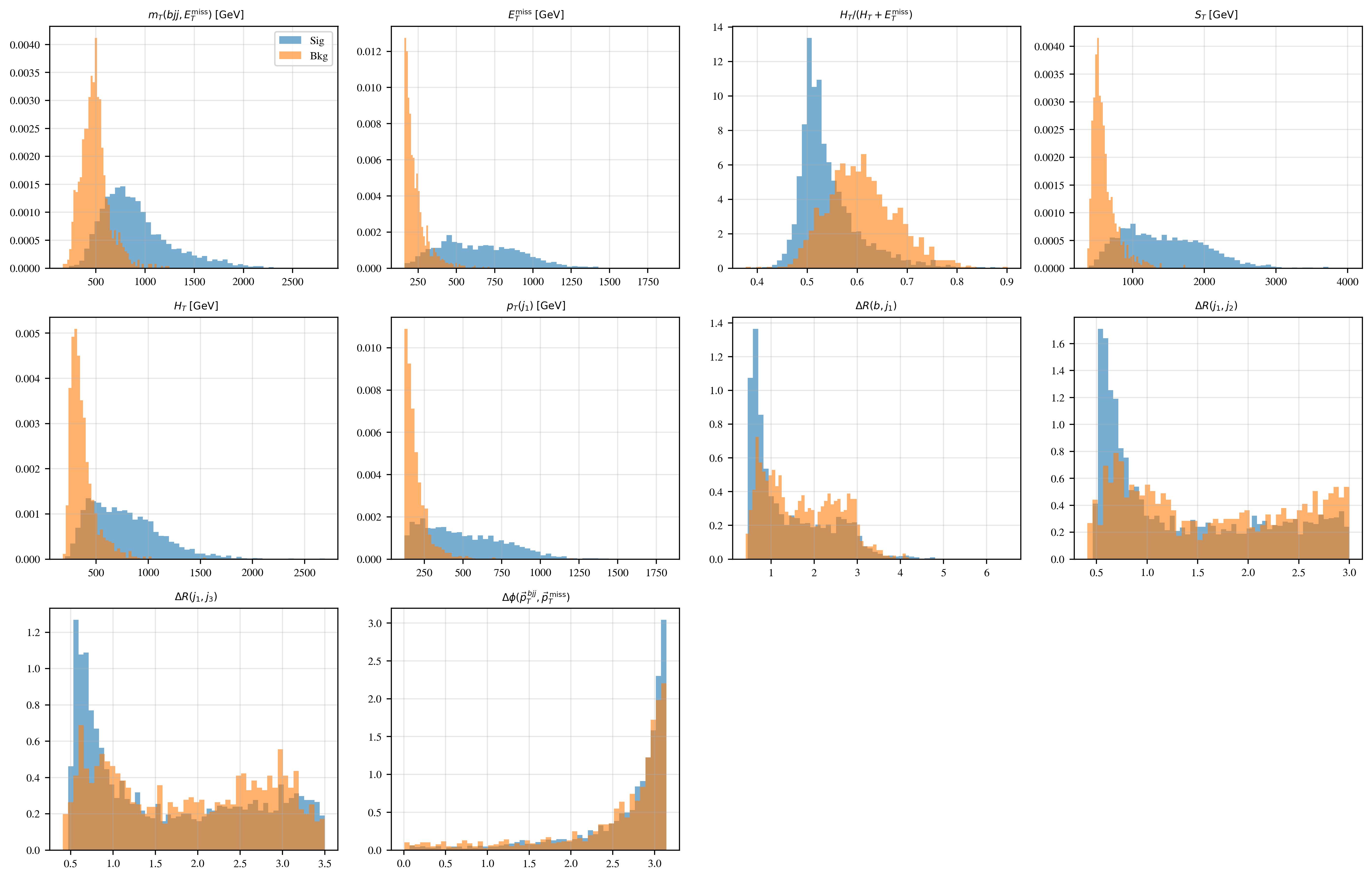}%
  }{%
    \fbox{\rule{0pt}{5cm}\rule{\dimexpr0.95\linewidth-2\fboxsep-2\fboxrule\relax}{0pt}}%
  }
\caption{Input variable distributions for GNN at the benchmark mass \(m_T=2000~\mathrm{GeV}\).}
  \label{fig:inputs_gnn}
\end{figure}

\begin{figure}[!htbp]
  \centering
  \begin{subfigure}[t]{0.49\linewidth}
    \centering
    \IfFileExists{f5_a.png}{%
      \includegraphics[width=\linewidth]{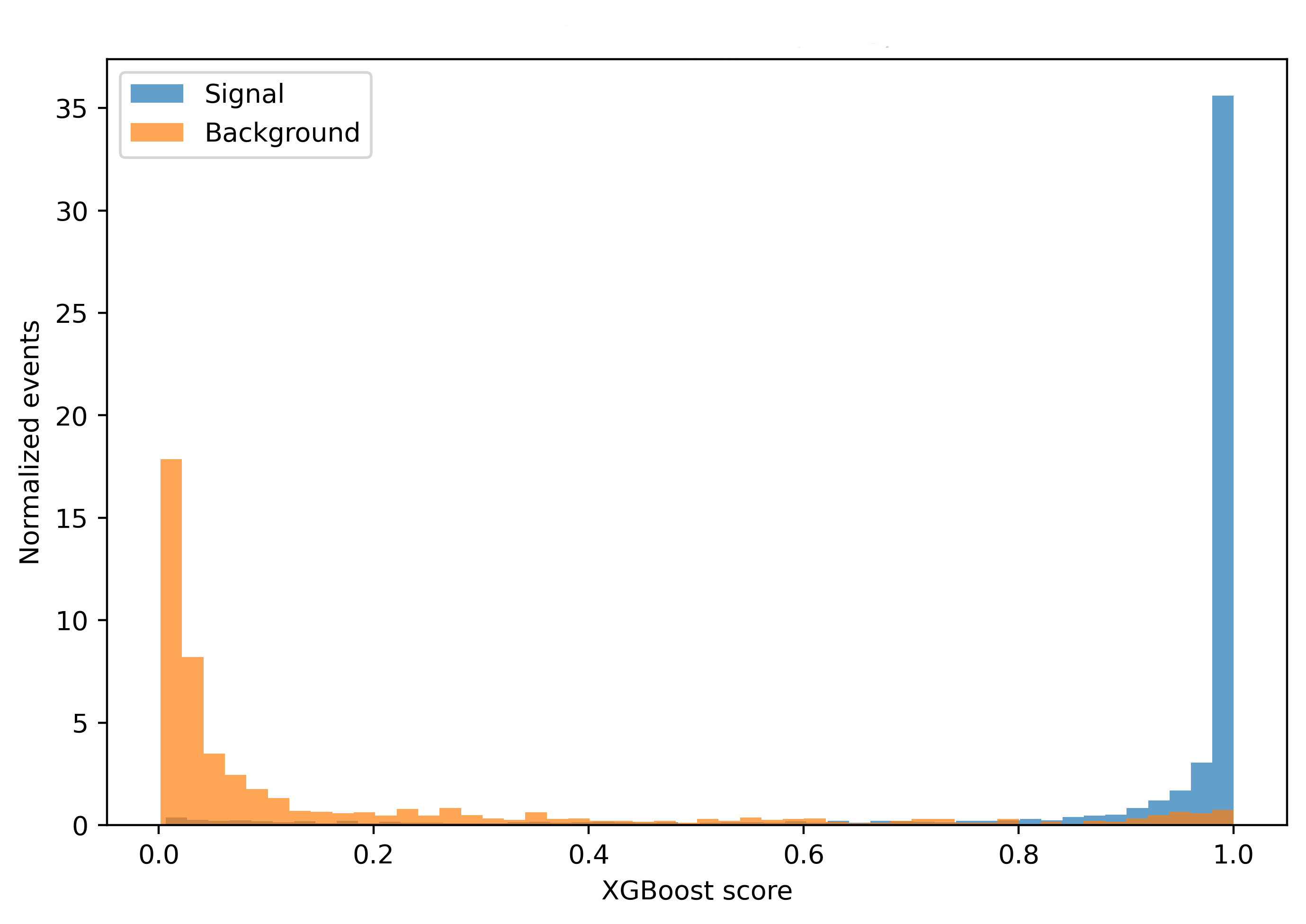}%
    }{%
      \fbox{\rule{0pt}{4cm}\rule{\dimexpr\linewidth-2\fboxsep-2\fboxrule\relax}{0pt}}%
    }
    \caption{XGBoost score distribution.}
  \end{subfigure}\hfill
  \begin{subfigure}[t]{0.49\linewidth}
    \centering
    \IfFileExists{f5_b.png}{%
      \includegraphics[width=\linewidth]{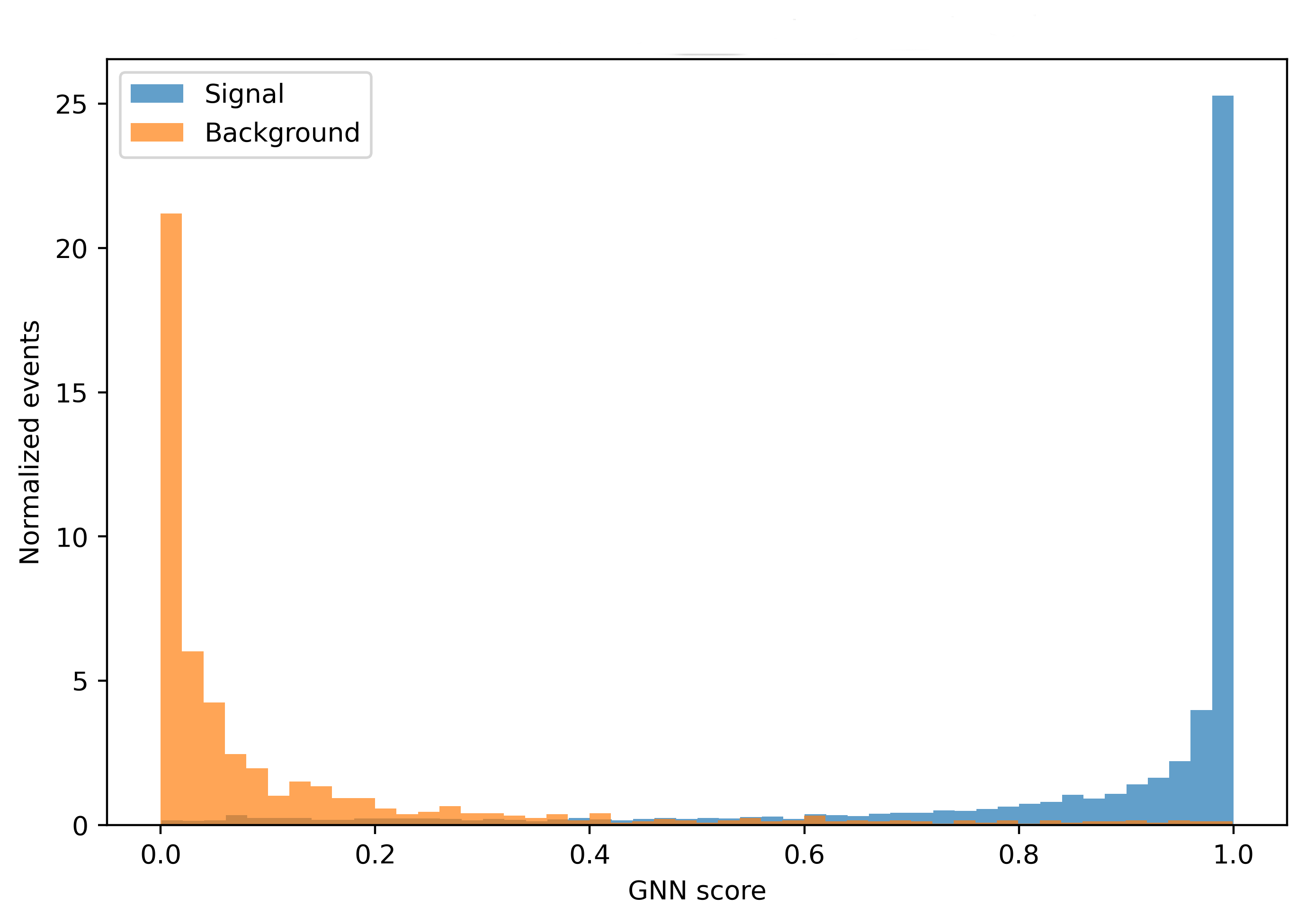}%
    }{%
      \fbox{\rule{0pt}{4cm}\rule{\dimexpr\linewidth-2\fboxsep-2\fboxrule\relax}{0pt}}%
    }
    \caption{GNN score distribution.}
  \end{subfigure}
  \caption{Recorded classifier score shapes for signal and background at \(m_T=2000~\mathrm{GeV}\) for (a) XGBoost and (b) GNN. The ordinate is labeled \enquote{normalized events} in the stored images, but values above unity indicate that it is likely a probability density rather than a bin fraction; the binning and normalization cannot be recovered.}
  \label{fig:score_dist}
\end{figure}

\begin{figure}[!htbp]
  \centering
  \begin{subfigure}[t]{0.49\linewidth}
    \centering
    \IfFileExists{f6_a.png}{%
      \includegraphics[width=\linewidth]{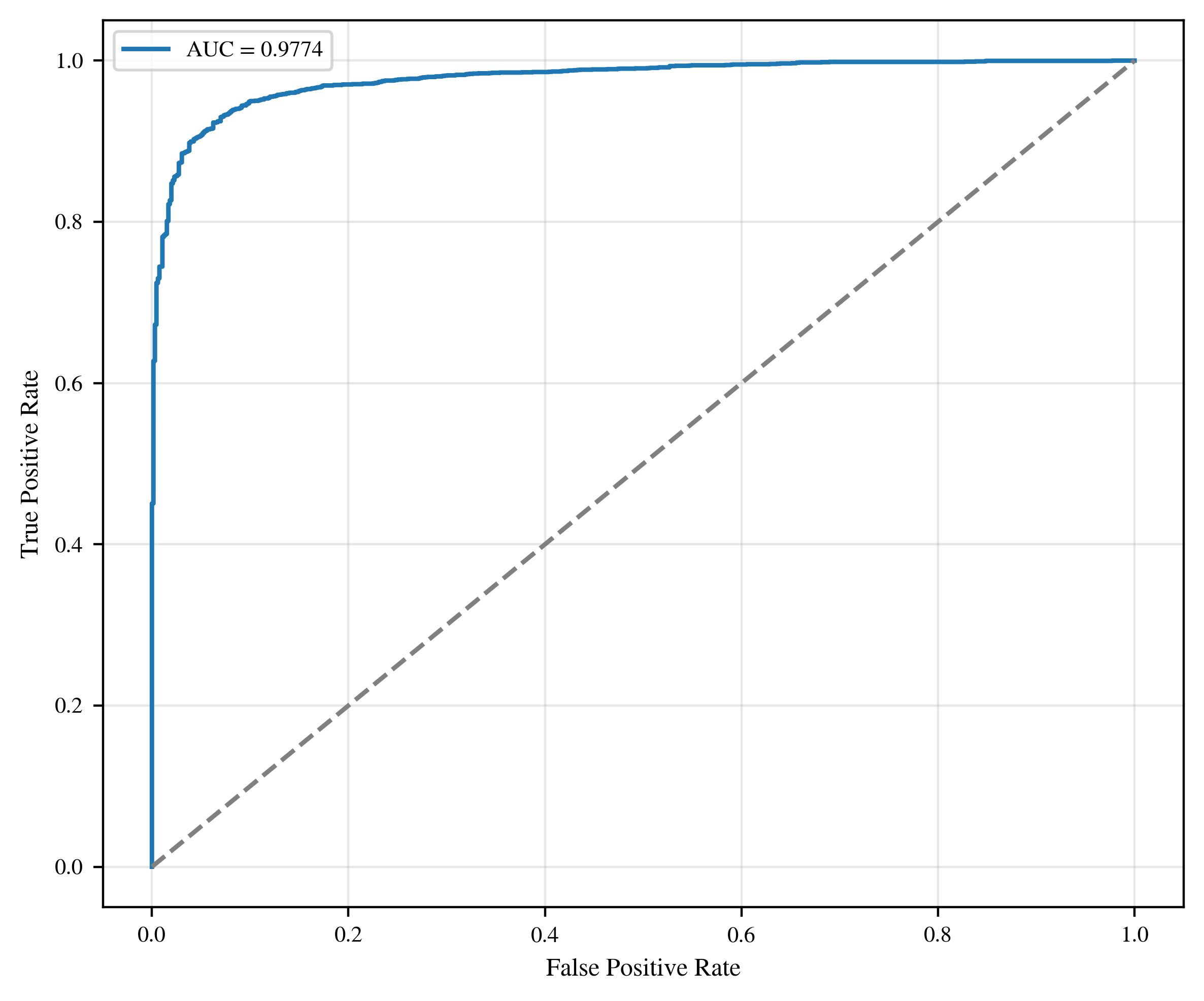}%
    }{%
      \fbox{\rule{0pt}{4cm}\rule{\dimexpr\linewidth-2\fboxsep-2\fboxrule\relax}{0pt}}%
    }
\caption{XGBoost ROC curve.}
  \end{subfigure}\hfill
  \begin{subfigure}[t]{0.49\linewidth}
    \centering
    \IfFileExists{f6_b.png}{%
      \includegraphics[width=\linewidth]{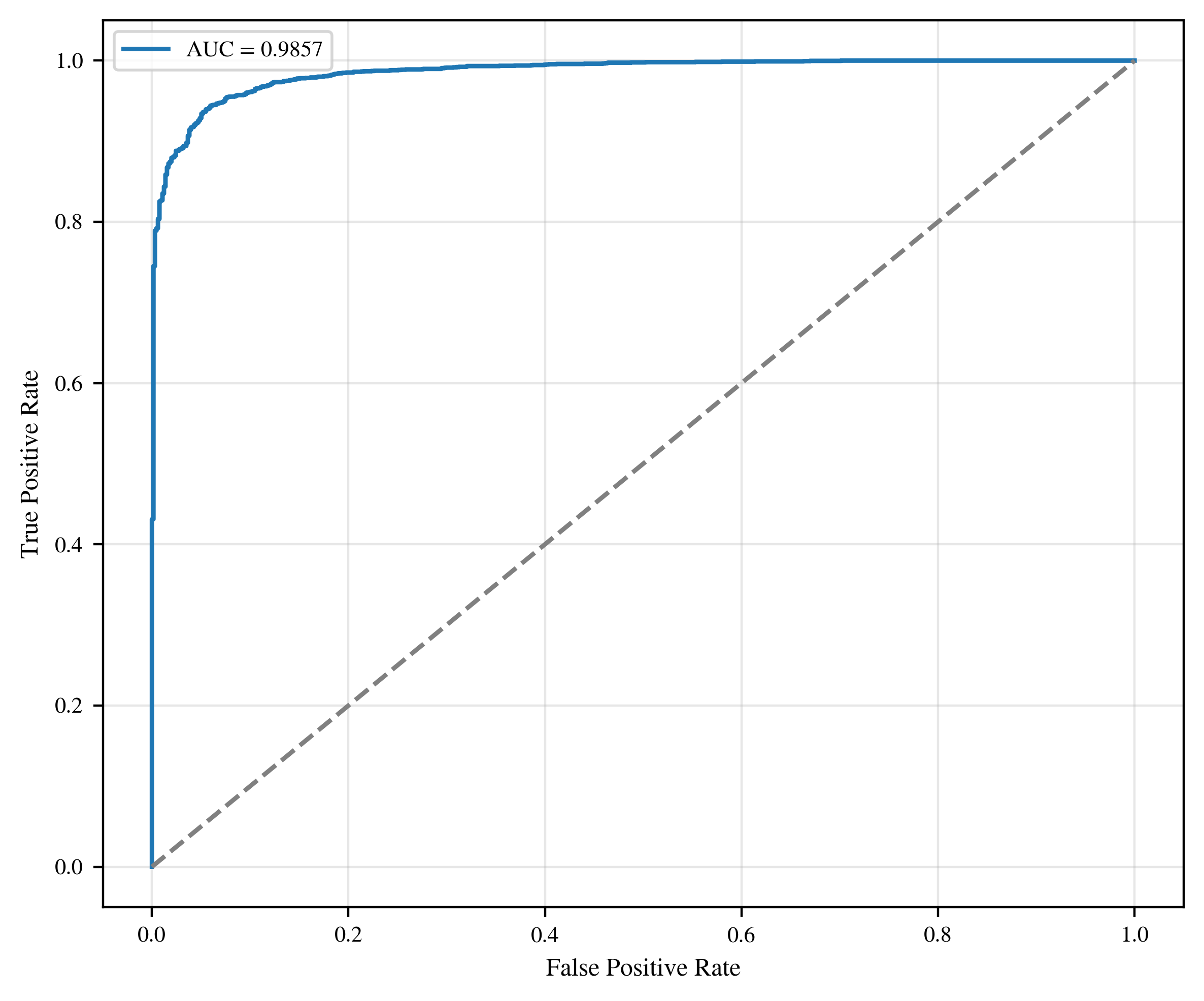}%
    }{%
      \fbox{\rule{0pt}{4cm}\rule{\dimexpr\linewidth-2\fboxsep-2\fboxrule\relax}{0pt}}%
    }
\caption{GNN ROC curve.}
  \end{subfigure}
\caption{ROC curves at $m_T=2000~\mathrm{GeV}$ for (a) XGBoost and (b) GNN.}
  \label{fig:roc}
\end{figure}

\rev{\Cref{fig:score_dist} shows the recorded classifier-score distributions, and \cref{fig:roc} shows the corresponding ROC curves. Across the mass scan, the test AUC ranges are 0.976--0.983 for XGBoost and 0.983--0.990 for the GNN, with a mean difference of \(7\times10^{-3}\). The GNN uses per-jet and pairwise information, whereas XGBoost uses engineered observables. These values therefore characterize two distinct pipelines and do not demonstrate the intrinsic superiority of either architecture; such a conclusion would require matched inputs and a common evaluation protocol.} The recorded KS \(p\)-values span 0.121--0.834 (signal) and 0.098--0.443 (background) for XGBoost, and 0.316--0.875 (signal) and 0.820--0.968 (background) for the GNN (\Cref{tab:ks_test}). None is below 0.05, but these single-split tests are insufficient to establish generalization in the absence of repeated seeds and independent samples.
\rev{The mild nonmonotonic variation of the historical GNN AUC near \(m_T=2.9\)~TeV may arise from limited sample size or training stochasticity, but the surviving outputs do not permit this hypothesis to be tested. The repeated-seed study in \cref{new:sec:analysis} provides the corresponding validation for the controlled common-input benchmark.}

\begin{table}[!htbp]
\centering
\small
\caption{Recorded KS test $p$-values comparing training and test score distributions for XGBoost and GNN.}
\label{tab:ks_test}
\begin{tabular}{|c|c|c|c|c|}
\hline
\multirow{2}{*}{$m_T$ [GeV]} & \multicolumn{2}{c}{XGBoost} & \multicolumn{2}{c}{GNN} \\
\cline{2-5}
 & $p_{\mathrm{KS}}^{\mathrm{sig}}$ & $p_{\mathrm{KS}}^{\mathrm{bkg}}$ & $p_{\mathrm{KS}}^{\mathrm{sig}}$ & $p_{\mathrm{KS}}^{\mathrm{bkg}}$ \\
\hline
1800 & 0.792 & 0.098 & 0.875 & 0.968 \\
2000 & 0.348 & 0.247 & 0.316 & 0.938 \\
2200 & 0.121 & 0.187 & 0.500 & 0.905 \\
2500 & 0.834 & 0.162 & 0.761 & 0.931 \\
2700 & 0.490 & 0.443 & 0.467 & 0.887 \\
2900 & 0.288 & 0.384 & 0.839 & 0.874 \\
3100 & 0.147 & 0.320 & 0.778 & 0.820 \\
\hline
\end{tabular}
\end{table}

\begin{figure}[!htbp]
  \centering
  \IfFileExists{f7.png}{%
    \includegraphics[width=0.58\linewidth]{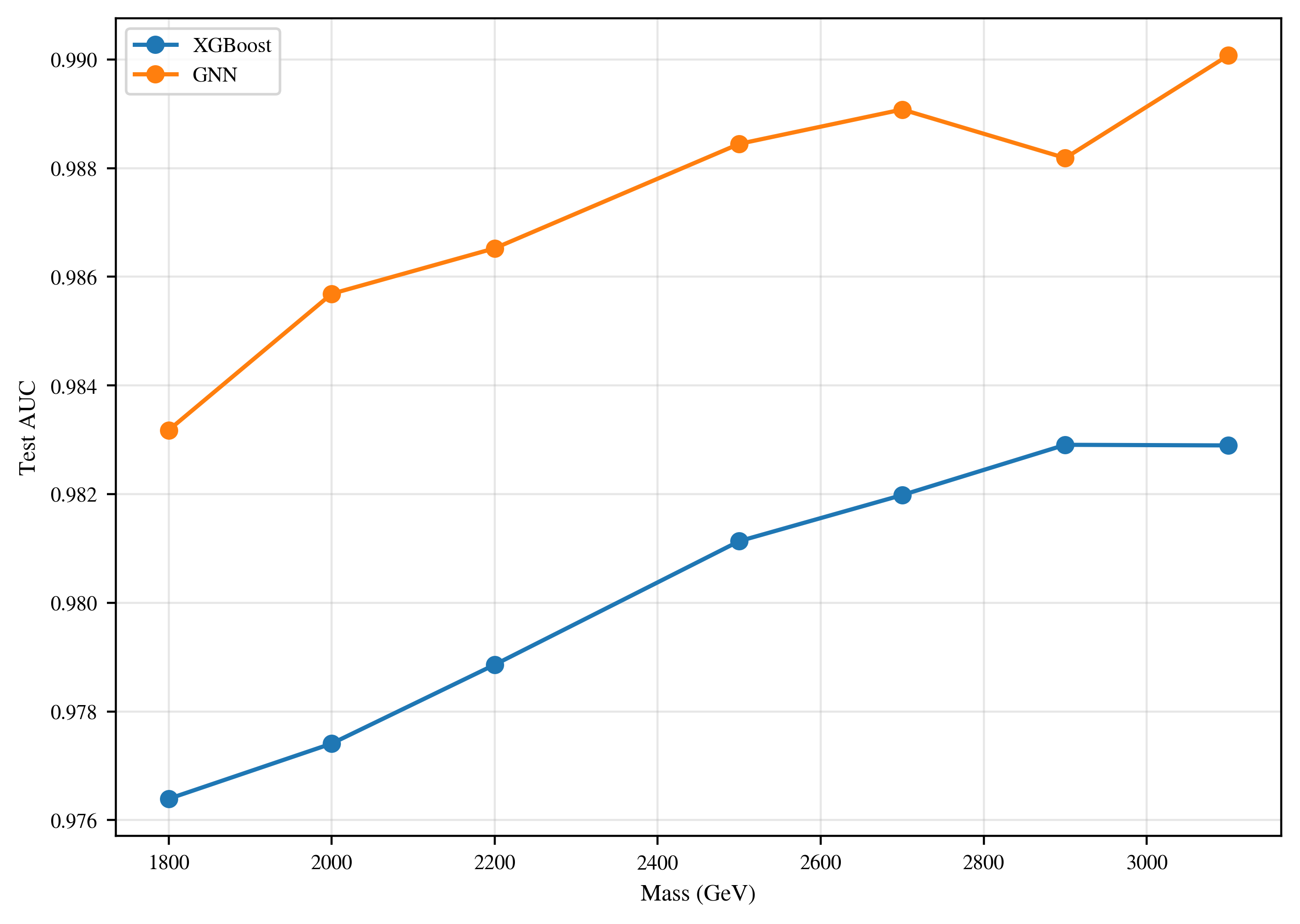}%
  }{%
    \fbox{\rule{0pt}{4cm}\rule{\dimexpr0.78\linewidth-2\fboxsep-2\fboxrule\relax}{0pt}}%
  }
\caption{{Recorded test AUC versus $m_T$ for the independently configured XGBoost and GNN pipelines. The curves are shown side by side descriptively; their nonidentical inputs preclude an architecture-level interpretation of the separation.}}
  \label{fig:auc_mass}
\end{figure}
\begin{figure}[!htbp]
  \centering
  \begin{subfigure}[t]{0.49\linewidth}
    \centering
    \IfFileExists{f8.png}{%
      \includegraphics[width=\linewidth]{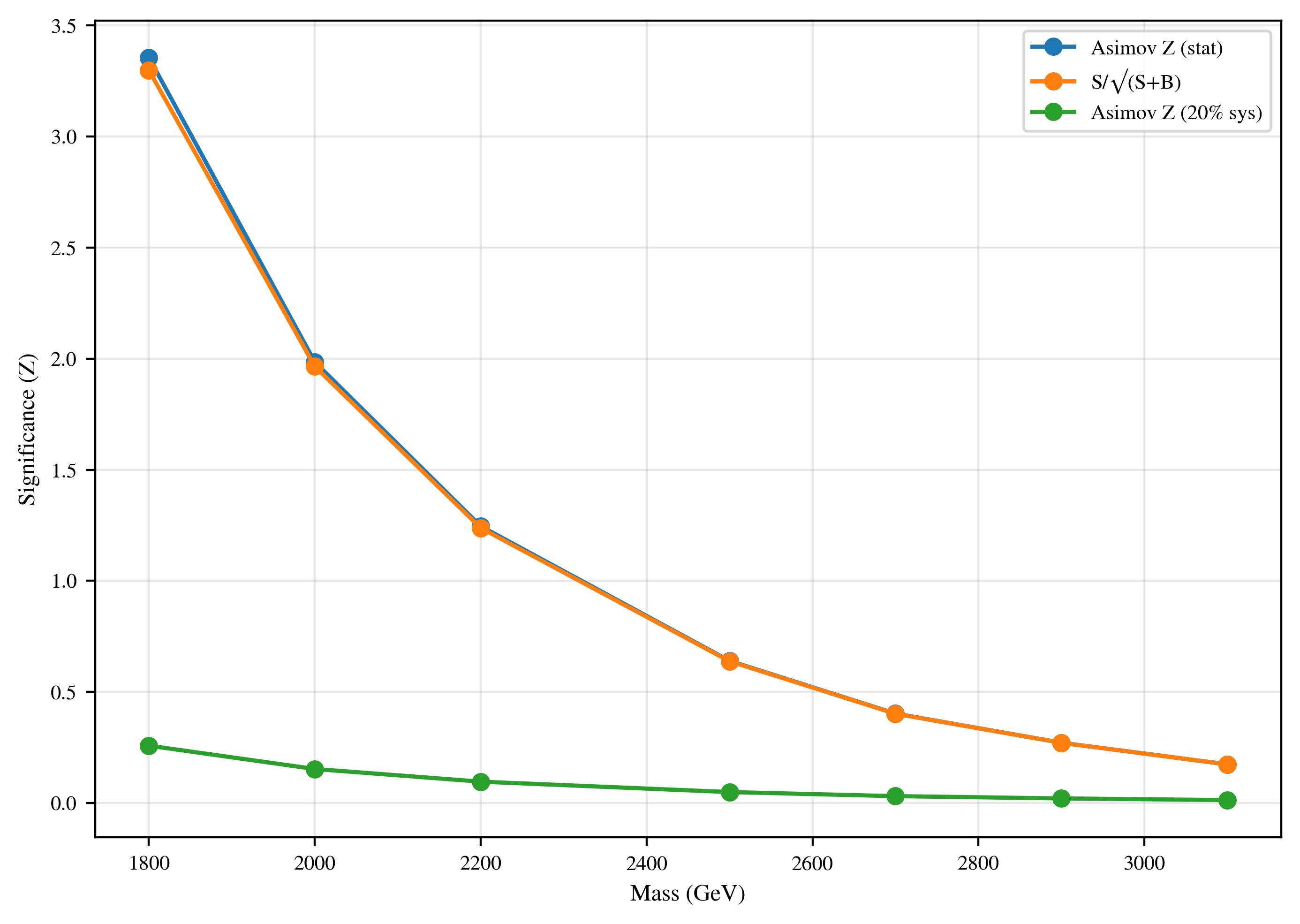}%
    }{%
      \fbox{\rule{0pt}{4cm}\rule{\dimexpr\linewidth-2\fboxsep-2\fboxrule\relax}{0pt}}%
    }
\caption{XGBoost significance.}
  \end{subfigure}\hfill
  \begin{subfigure}[t]{0.49\linewidth}
    \centering
    \IfFileExists{f9.png}{%
      \includegraphics[width=\linewidth]{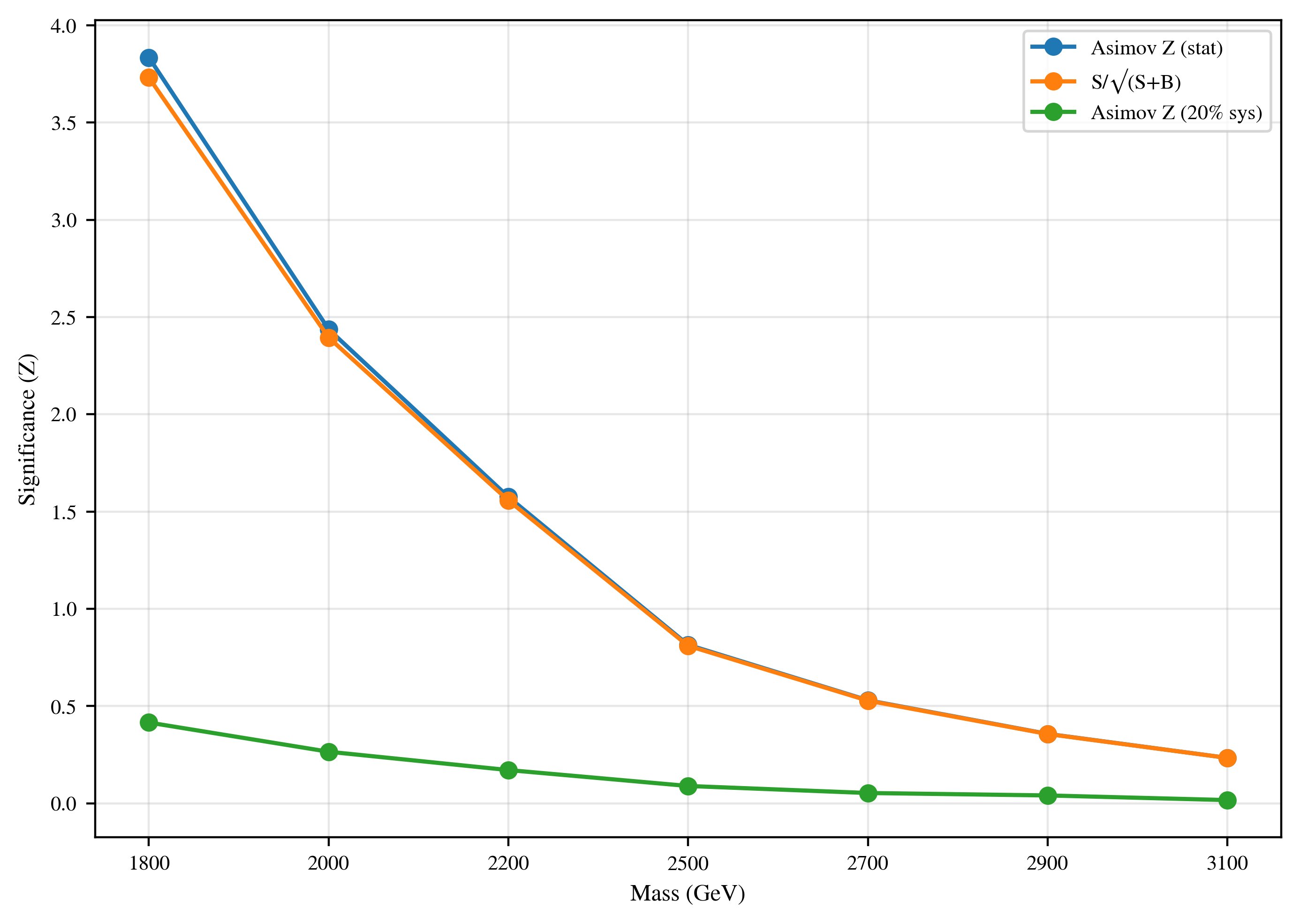}%
    }{%
      \fbox{\rule{0pt}{4cm}\rule{\dimexpr\linewidth-2\fboxsep-2\fboxrule\relax}{0pt}}%
    }
\caption{GNN significance.}
  \end{subfigure}
\caption{{Recorded significance measures vs. \(m_T\) at the historical threshold \(s_0=0.99\) and the nominal sample label \(g^*=0.4,\ R_L=0.5\), for (a) XGBoost and (b) GNN. The common numerical score threshold is retained only for these historical sensitivity outputs and is not interpreted as an equivalent classifier operating point; the direct event-level fixed-\(\epsilon_S\) and validation-selected comparisons are given in \cref{new:tab:fixed,new:tab:wp}. Their historical luminosity normalization is audited in \cref{new:sec:historical}.}}
  \label{fig:z_mass}
\end{figure}

% CPC-2026-0641 normalization revision: original contour figure retained in source but excluded from compiled manuscript.
\iffalse
\begin{figure}[!htbp]
  \centering
  \begin{subfigure}[t]{0.49\linewidth}
    \centering
    \IfFileExists{f10.png}{%
      \includegraphics[width=\linewidth]{f10.png}%
    }{%
      \fbox{\rule{0pt}{4cm}\rule{\dimexpr\linewidth-2\fboxsep-2\fboxrule\relax}{0pt}}%
    }
    \caption{Original label \(R_L=0\).}
    \label{fig:contours_rl0}
  \end{subfigure}\hfill
  \begin{subfigure}[t]{0.49\linewidth}
    \centering
    \IfFileExists{f11.png}{%
      \includegraphics[width=\linewidth]{f11.png}%
    }{%
      \fbox{\rule{0pt}{4cm}\rule{\dimexpr\linewidth-2\fboxsep-2\fboxrule\relax}{0pt}}%
    }
    \caption{Original label \(R_L=0.5\).}
    \label{fig:contours_rl05}
  \end{subfigure}
  \caption{Original statistical-only \(Z_A=2\) and \(Z_A=5\) diagnostic contours in the \((g^*,m_T)\) plane. The missing \(R_L\)-dependent production-rate reweighting prevents a physical interpretation, these are not profile-likelihood confidence contours, and the meaning of the shaded regions is not documented.}
\end{figure}
\fi

\rev{The stored sensitivity results use \(s_0=0.99\) for both historical classifiers. Because this numerical threshold does not define equivalent operating points, it is not used to compare the classifiers; the controlled event-level comparison is presented in \cref{new:tab:fixed,new:tab:wp}.} \Cref{fig:z_mass} shows the three recorded historical sensitivity measures versus \(m_T\). For XGBoost, the ranges are \(Z_A=0.174\)--3.355, \(S/\sqrt{S+B}=0.174\)--3.298, and \(Z_{A,\sigma_B}=0.0135\)--0.258. For the GNN they are \(Z_A=0.234\)--3.835, \(S/\sqrt{S+B}=0.234\)--3.731, and \(Z_{A,\sigma_B}=0.017\)--0.416. Thus no stored mass point reaches \(Z=2\) once a 20\% background-normalization uncertainty is included. \rev{Here, \(B\) is the sum of the four available background samples. The 20\% uncertainty is applied to this total yield and does not account for unsimulated background processes. The earlier raster-derived operating-point table and significance-inversion estimates are omitted from the revised quantitative results and superseded by the direct event-level results in \cref{new:tab:fixed,new:tab:wp,new:sec:historical}.} \rev{Consequently, no \(g^*\)-dependent \(Z_A=2\) or \(Z_A=5\) contours are inferred from these benchmark values.}
\FloatBarrier

\FloatBarrier

% Original diagnostic discussion retained for traceability but excluded from the compiled revised manuscript.
\iffalse
There is also an independent normalization contradiction. The mass-scan samples and \cref{fig:z_mass} are labeled \(g^*=0.4,\ R_L=0.5\), yet their largest statistical significance is only \(Z_A=3.835\). Under the stated positive quadratic signal-rate rescaling, reducing \(g^*\) cannot increase \(S\) or \(Z_A\). Nevertheless, the table in \hyperref[sec:appendix]{Appendix~\ref*{sec:appendix}} (\cref{tab:gstar_thresholds}) records \(Z_A=5\) crossings at \(g^*=0.31\)--0.49 for the same \(R_L=0.5\) label. Therefore the threshold table cannot have been derived consistently from the displayed benchmark scan; this contradiction alone prevents interpreting the contours as physical reach.

\Cref{fig:contours_rl0,fig:contours_rl05} retain the original rate contours. Internally, these plots require larger \(g^*\) at \(R_L=0.5\) than at \(R_L=0\), whereas the production calculation in Ref.~\cite{LiChaoZhang2023} finds that first-generation mixing enhances single production through valence-quark luminosities. The recorded trend is consistent with applying the decrease of \(\mathrm{Br}(T\to tZ)\) while omitting the compensating \(R_L\)-dependence of \(\sigma(pp\to Tj)\). The tables and plots are therefore diagnostics of the stored calculation, not physical comparisons of the two mixing scenarios.
\fi

\rev{Before revision, the original coupling-rescaling results were compared with the directly stored benchmark scan. This check revealed a normalization inconsistency: the samples are labeled \(g^*=0.4\) and \(R_L=0.5\), whereas the original threshold results imply five-standard-deviation crossings at smaller \(g^*\), contrary to a positive quadratic signal-rate scaling. The original \((g^*,m_T)\) contours and coupling-threshold table are therefore excluded from the revised quantitative results. All subsequent sensitivity statements refer only to the directly simulated benchmark at \(g^*=0.4\) and \(R_L=0.5\); no coupling-dependent exclusion or discovery reach is claimed.}

{\subsection{Controlled machine-learning comparison}\label{new:sec:analysis}
The stored feature tables, cutflow reports and event weights can now be inspected directly. \rev{The recovered normalization uses $K=1.15$ for $ZZjj$, confirming the correction described in \cref{sec:simulation}; the corresponding factors are 1.14 for $WZjj$, 1.10 for $tZj$, 1.616 for $t\bar t$, and 0.95 for the signal.} These factors are retained to reproduce the existing sample normalization, rather than asserted to be universal higher-order corrections \cite{Czakon2013ttbar,Fuks2017VLQNLO,Cacciapaglia2019Kfactor}. The 2 TeV feature table contains 15\,555 signal entries and 6\,114 background entries: 3\,353 $t\bar t$, 1\,416 $tZj$, 964 $ZZjj$ and 381 $WZjj$. All stored weights are positive. Their pre-classifier sums are 1007.93 signal events and $6.85998\times10^6$ background events under the original $3000~\mathrm{fb}^{-1}$ convention.

The recovered final 2 TeV sums agree with the last row of \cref{tab:cutflow_bench} to its printed precision. This checks stored-feature bookkeeping; upstream generator cards and reconstruction provenance remain incomplete.

\subsubsection{Common inputs and event partitions}
Both models receive 13 global quantities: the stored $m_T$ observable, $m_{bjj}$, $E_T^{\rm miss}$, $H_T$, $S_T$, $H_T/S_T$, $\Delta\phi(bjj,\vec p_T^{\,\rm miss})$, five recorded separations $\Delta R(j_1,j_2)$, $\Delta R(j_1,j_3)$, $\Delta R(j_2,j_3)$, $\Delta R(b,j_1)$, $\Delta R(b,j_2)$, and $p_T(j_1)$. In these files $H_T=p_T(j_1)+p_T(j_2)+p_T(j_3)$ and $S_T=H_T+E_T^{\rm miss}$; $H_T$ is not the scalar sum over all reconstructed jets.

The four node roles are the three leading inclusive jets and the stored $b$ candidate. The latter can coincide with one of the leading jets, so the graph does not necessarily contain four distinct physical jets. Each node supplies $\log(1+p_T/\mathrm{GeV})$, $\eta$, $\sin\phi$, $\cos\phi$, and $\log[1+\max(m,0)/\mathrm{GeV}]$, together with a four-component one-hot role label. Every ordered pair of different roles is connected by an edge whose feature is $\Delta R$, calculated directly from these node coordinates with wrapped azimuth. The 12 directed edge values are supplied to both classifiers, even though opposite directions are redundant. These geometric edges are distinct from the stored reconstruction separations retained among the global inputs. XGBoost receives the flattened 61-component array; the GNN receives the identical node, edge and global tensors. No process identifiers, event indices or weights enter the prediction features.

Each physical process is independently divided into approximately 60\% training, 20\% validation and 20\% test events with split seed 42. The partitions contain 12\,999, 4\,334 and 4\,336 events respectively. Exact comparisons of stored kinematic records find no duplicates across the three partitions. This is a check of stored-record identity, not proof about unrecorded generator ancestry. The split is shared by all models and all training seeds; each model is trained separately at one mass. The repeated background records have identical split assignments at every mass, as verified from their stored identities; none crosses a partition boundary between masses. Since the same backgrounds are reused, mass-point results are correlated.

Continuous node and global features are standardized using training-only moments, and edge distances use a training-only common scale. The training signal weight is multiplied by the training background-to-signal weight-sum ratio, after which all training weights share one normalization to unit mean. The same transformation is used by both classifiers and on validation loss. Physical weights, without class balancing, define the ROC curves and AUC. Class balancing means that a score near unity is not a calibrated signal probability in the physical event mixture.

\subsubsection{Models and stopping rules}
XGBoost uses at most 350 depth-three trees, learning rate 0.05, row and feature subsampling fractions 0.7, minimum child weight 2, split penalty 0.2, $L_2$ penalty 1 and zero $L_1$ penalty. Training stops after 20 rounds without improvement in validation log loss; predictions use the best iteration. These settings follow the existing tree configuration, with an explicit validation stopping rule \cite{Chen2016XGBoost}.

The new GNN embeds each nine-component node in 32 dimensions. In each of two message-passing layers, a linear map followed by ReLU acts on the concatenated receiver embedding, sender embedding and edge distance. The three incoming messages per role are averaged. A second linear map of the node embedding and this average is followed by ReLU and dropout 0.2. Mean pooling over the four roles gives a graph vector. Concatenating the 13 global quantities, a 64-unit ReLU layer with dropout 0.2 and a final sigmoid gives the score. Role labels preserve the leading-jet ordering information also supplied to the flattened classifier. This is a newly specified message-passing benchmark, not a reproduction of the old NNConv checkpoint, which was not saved.

The GNN uses Adam with learning rate $10^{-3}$, weight decay $10^{-4}$, batch size 512 and at most 60 epochs. Weighted binary cross entropy is averaged using the fixed training-derived weight normalization. The checkpoint is selected by validation log loss, with patience ten epochs and minimum improvement $10^{-4}$. No hyperparameter search is performed. Model seeds 42, 137 and 2026 test training variability on the fixed event partition. Seed 42 was designated for the detailed figures before training. \rev{The analysis uses XGBoost 3.1.2 and PyTorch 2.9.1 on a CPU. The retained artifacts include preprocessing constants, model files, split identifiers, training histories, and event-level predictions and are available from the authors on reasonable request.}

\subsubsection{Efficiencies, yields and working points}
We compare background rejection at empirical test-sample signal efficiencies of 50\%, 70\% and 80\%. These ROC readouts characterize discrimination; they are not thresholds optimized on the test sample for a sensitivity claim. Separately, a threshold is selected on validation events for each target efficiency and then frozen before evaluation on the test set. The first attainable weighted efficiency at or above each target is used, with the acceptance convention $s\ge s_0$ throughout.

For a process $p$ and partition $D$, the stored weights are expanded by $N_p/N_{p,D}$, where $N_p$ counts all stored pre-classifier events and $N_{p,D}$ counts that partition. Each process has constant stored weights, so this is also the ratio of its full to partition weight sums. Thus
\begin{equation}
Y_{p,D}(s_0)=\frac{N_p}{N_{p,D}}\sum_{i\in(p,D)}w_i\,\mathbf1(s_i\ge s_0),\qquad
\epsilon_{p,D}=\frac{Y_{p,D}(s_0)}{Y_{p,D}(-\infty)}.
\label{new:eq:expanded}
\end{equation}
Summing signal or background processes gives $S$, $B$, $\epsilon_S$ and $\epsilon_B$. The stored preselection normalization is held fixed. We also give raw accepted counts and the weighted-simulation diagnostic
\begin{equation}
N_{\rm eff,B}=\frac{(\sum_{i\in B}w_i')^2}{\sum_{i\in B}(w_i')^2},\qquad w_i'=w_i N_p/N_{p,D}.
\end{equation}
The reported $\sqrt{\sum_i(w_i')^2}$ is the usual Monte Carlo counting error estimate, conditional on the assigned cross-section weights. It excludes normalization, detector and modeling uncertainties. A zero observed count in a process is not evidence that its physical yield vanishes.

As a separate diagnostic, we select a validation threshold maximizing the counting Asimov expression with $\sigma_B=0.2B$, subject to at least ten accepted signal events and $N_{\rm eff,B}\ge10$. The formula follows the counting-likelihood treatment of Ref.~\cite{Cowan2011Asimov}. Explicitly, the diagnostic objective is
\begin{equation}
Z_{20}^2=2\left[(S+B)\ln\frac{(S+B)(B+v)}{B^2+(S+B)v}-\frac{B^2}{v}\ln\left(1+\frac{vS}{B(B+v)}\right)\right],\quad v=(0.2B)^2.
\end{equation}
This minimum support criterion is a declared analysis choice, not a guarantee of precise tail modeling. The chosen thresholds and their independent test support are reported; they are not retuned if the test support is lower. No discovery significance is inferred from the sparse fixed-efficiency tails, and a 20\% normalization nuisance cannot account for omitted processes.
\subsection{Results and validation}\label{new:sec:results}
\Cref{new:tab:auc} gives the AUC for every training seed. XGBoost has the larger test AUC in all three runs, with a difference of 0.0009--0.0023. The spread across seeds measures training variability on one split; it is not a confidence interval for performance on independent simulated samples. No general superiority of either architecture follows from this restricted benchmark.
\begin{table}[htbp]\centering\small
\caption{Physical-weight AUC for the controlled 2 TeV comparison. Best step denotes a boosting round or GNN epoch.}\label{new:tab:auc}
\begin{tabular}{llrrrr}\hline
Seed & Model & Train AUC & Validation AUC & Test AUC & Best step \\
\hline
42 & XGBoost & 0.99386 & 0.98637 & 0.98623 & 195 \\
42 & GNN & 0.99018 & 0.98708 & 0.98449 & 21 \\
137 & XGBoost & 0.99485 & 0.98674 & 0.98655 & 235 \\
137 & GNN & 0.99018 & 0.98690 & 0.98428 & 22 \\
2026 & XGBoost & 0.99359 & 0.98639 & 0.98562 & 172 \\
2026 & GNN & 0.99035 & 0.98694 & 0.98472 & 25 \\
\hline\end{tabular}
\end{table}
Separate physical-weight Pearson matrices for signal and the summed background are shown in \cref{new:fig:corrs,new:fig:corrb}. They use all stored events after the pre-classifier selection and are descriptive diagnostics, not inputs to feature selection. In particular, $S_T=H_T+E_T^{\rm miss}$ and the balance ratio introduce deterministic nonlinear redundancy. Such correlations do not themselves constitute leakage. \rev{The complete correlation matrices for all nonconstant inputs, including the node and geometric-edge variables, are shown in \cref{new:fig:allcorrsignal,new:fig:allcorrbackground}.}
\begin{figure}[H]\centering\includegraphics[width=.95\linewidth]{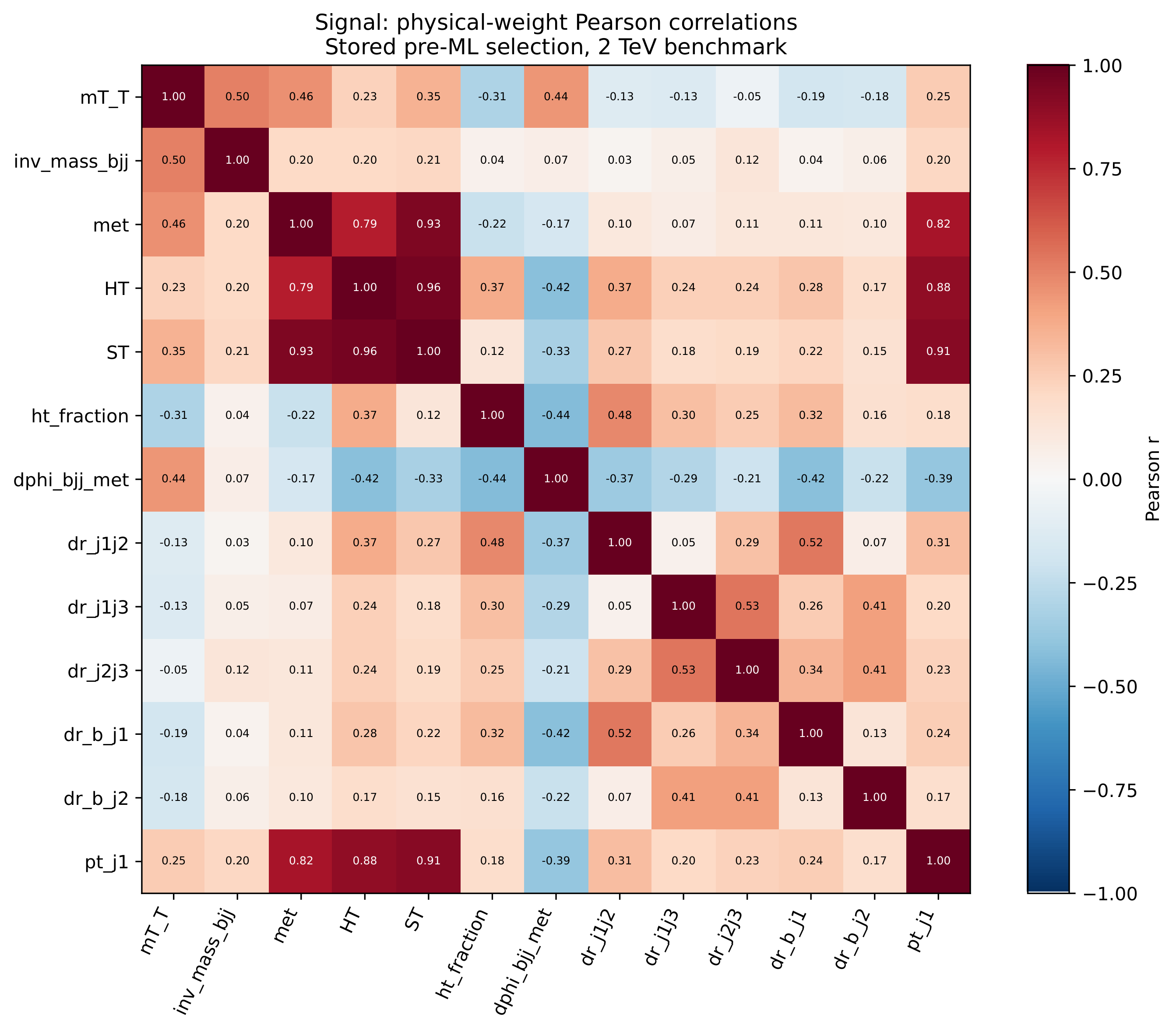}
\caption{Global-input Pearson correlations for the stored 2 TeV signal population.}\label{new:fig:corrs}
\end{figure}
\begin{figure}[H]\centering\includegraphics[width=.95\linewidth]{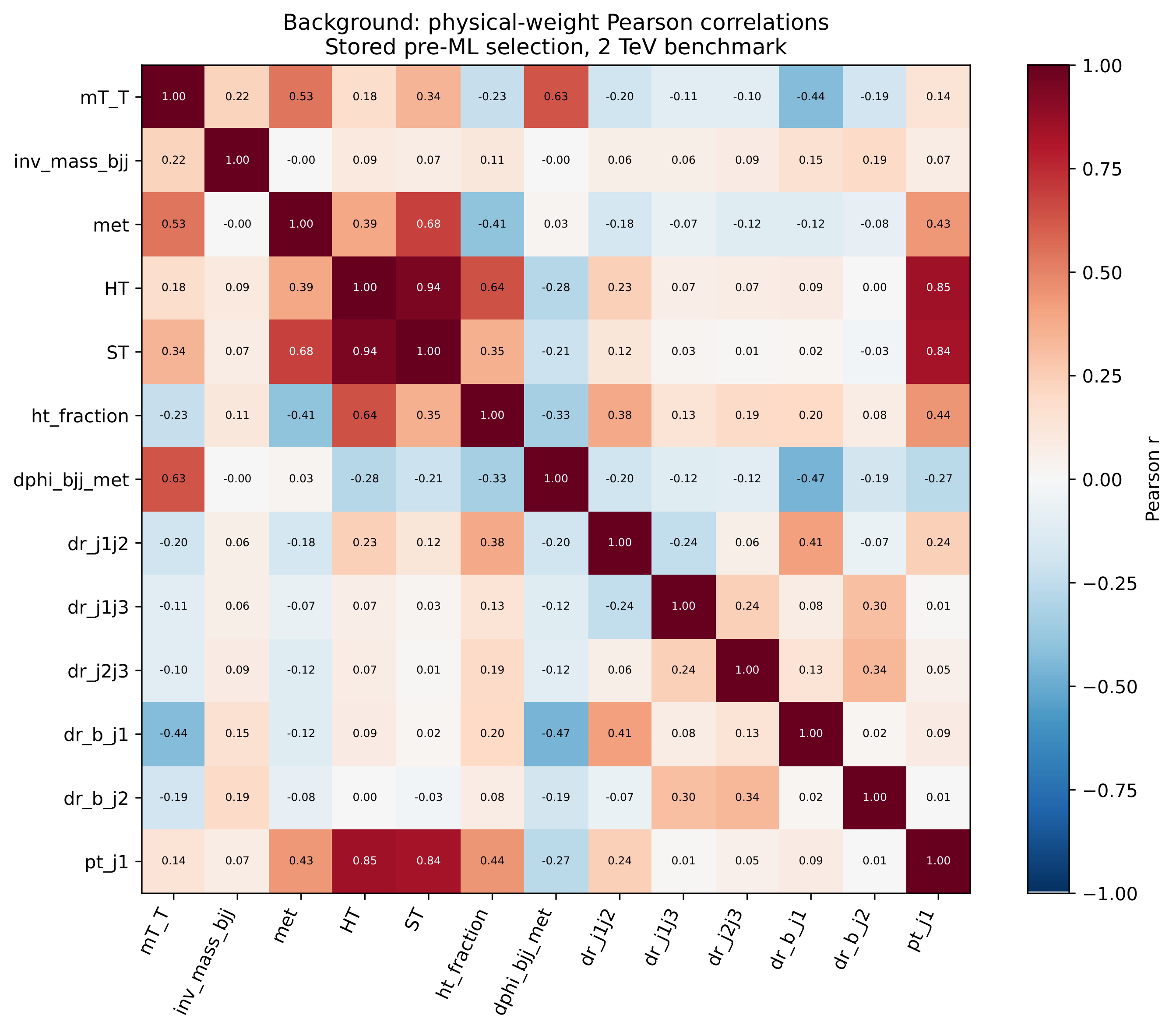}
\caption{Global-input Pearson correlations for the stored four-process background mixture, using physical event weights.}\label{new:fig:corrb}
\end{figure}
\Cref{new:fig:learning} shows the evaluated training and validation loss and AUC at every step for seed 42. The selected checkpoints are XGBoost round 195 and GNN epoch 21. The training--validation gap is visible and is not dismissed by the large AUC. The separate signal and background train/test score densities in \cref{new:fig:scores} are normalized within class and partition, with a logarithmic ordinate. These distributions are based on saved event-level scores and retain explicit bin boundaries at 0.99.

Unweighted two-sample KS diagnostics for seed 42 give $(D,p)=(0.0132,0.807)$ for XGBoost signal, $(0.0438,0.0578)$ for XGBoost background, $(0.0158,0.603)$ for GNN signal and $(0.0333,0.254)$ for GNN background. These conventional $p$-values are only shape diagnostics: the training observations helped fit the estimator, and the unweighted background composition differs from the physical mixture. They neither certify independence nor exclude overfitting. \rev{Among the retained seed-level diagnostics, the XGBoost background sample gives a minimum value of $p=0.0180$ for seed 137. No seed is excluded on this basis, and the complete numerical results are available from the authors on reasonable request.}
\begin{figure}[H]\centering\includegraphics[width=.95\linewidth]{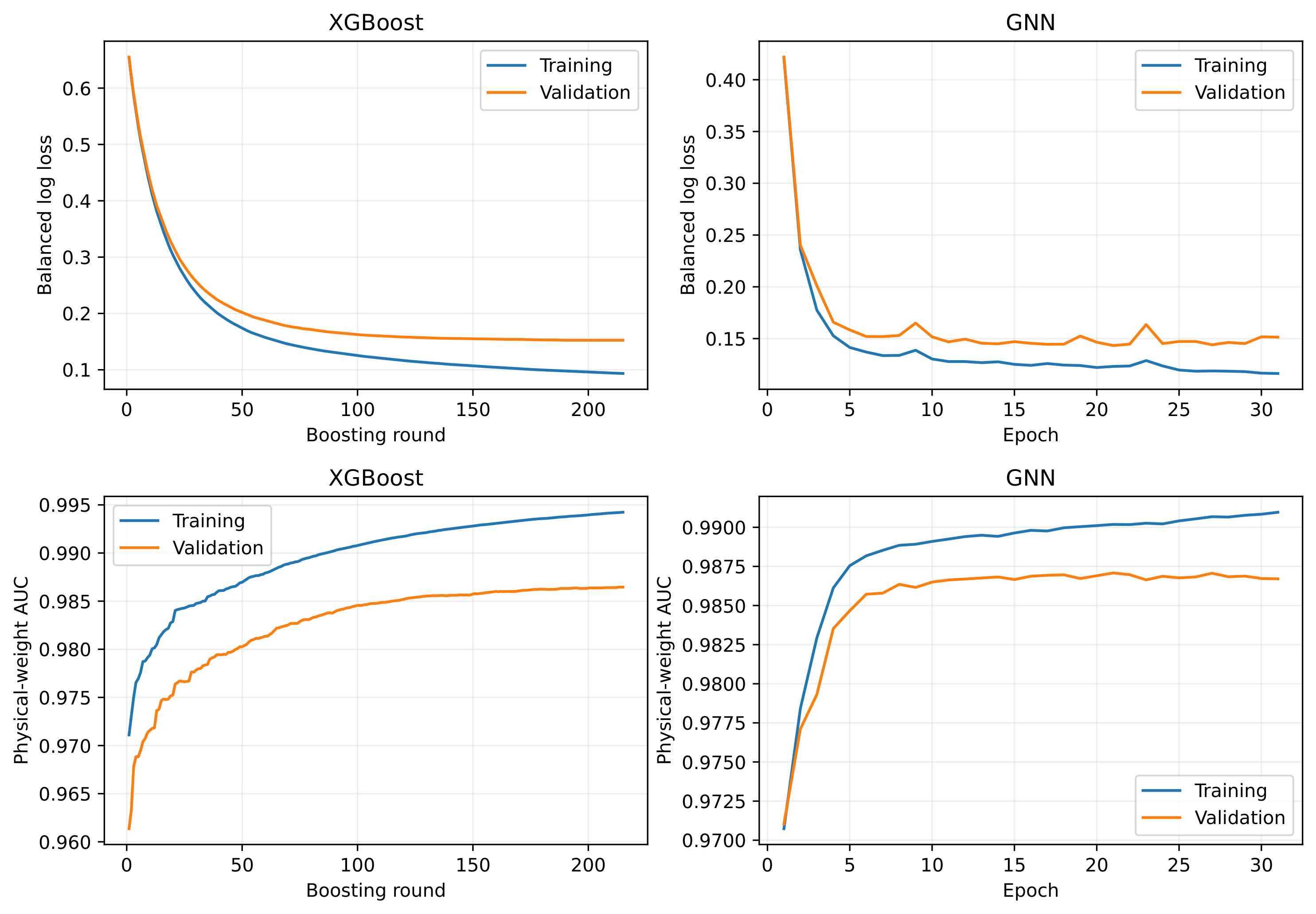}
\caption{Training and validation histories for the controlled \(m_T=2~\mathrm{TeV}\), seed-42 benchmark. Loss uses the common class-balanced weighting; AUC uses physical weights.}\label{new:fig:learning}
\end{figure}
\begin{figure}[H]\centering\includegraphics[width=.95\linewidth]{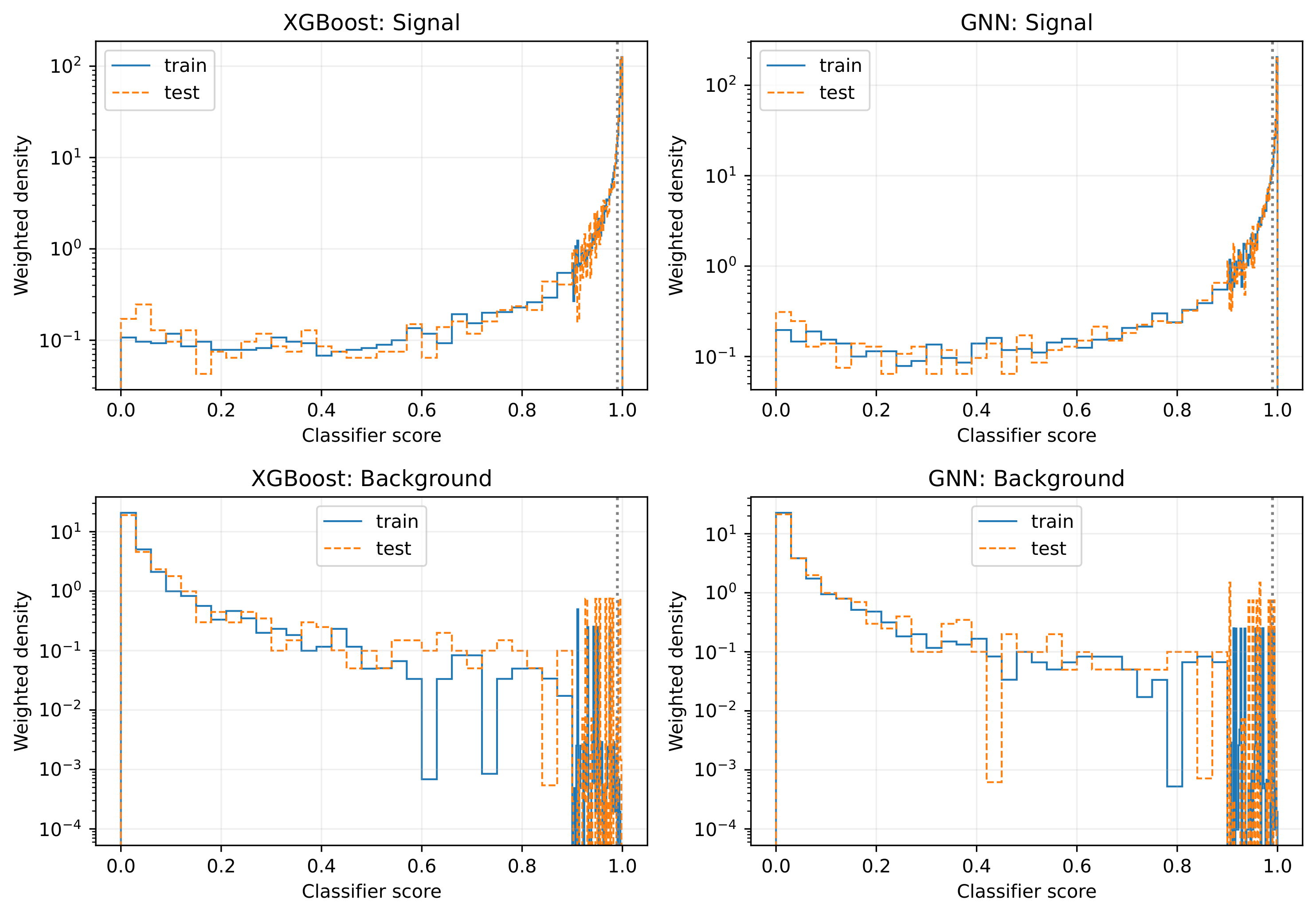}
\caption{Separate training and test score densities for the controlled \(m_T=2~\mathrm{TeV}\), seed-42 benchmark, normalized within each class and partition with physical weights. Empty bins have no logarithmic ordinate.}\label{new:fig:scores}
\end{figure}
\begin{figure}[H]\centering\includegraphics[width=.95\linewidth]{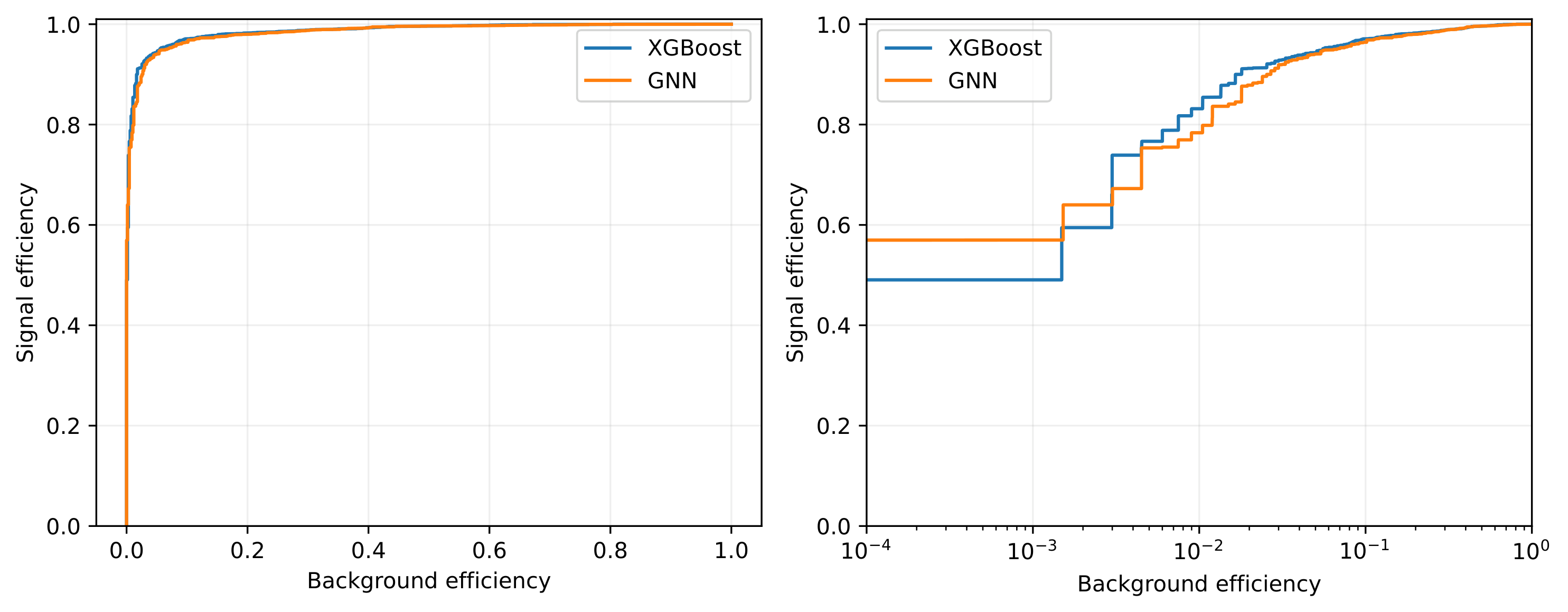}
\caption{Held-out ROC curves for the controlled \(m_T=2~\mathrm{TeV}\), seed-42 benchmark with linear and logarithmic background-efficiency axes. The log panel omits points at exactly zero observed background efficiency.}\label{new:fig:roc}
\end{figure}
\begin{table}[H]\centering\small
\caption{Empirical weighted test ROC readouts for the controlled \(m_T=2~\mathrm{TeV}\), seed-42 benchmark; the first attainable signal efficiencies are 50.016, 70.010 and 80.006 percent. These are descriptive ROC points, not validation-selected cuts. The near-perfect 50 percent GNN rejection has very little simulated background support.}\label{new:tab:fixed}
\begin{tabular}{rrr}\hline
Target $\epsilon_S$ (\%) & XGBoost rejection (\%) & GNN rejection (\%) \\
\hline
50 & 99.8510 & 99.9971 \\
70 & 99.7006 & 99.5502 \\
80 & 99.2510 & 98.8048 \\
\hline\end{tabular}
\end{table}
\Cref{new:tab:wp} answers a different question: how thresholds chosen solely on validation events perform on held-out events. Both classifiers receive the same target efficiencies, but their thresholds differ. The historical numerical cut 0.99 is also included as a diagnostic. For example, it retains 66.09\% of XGBoost signal and 60.88\% of GNN signal, so it is not an equivalent operating point.
\begin{table}[H]\centering\small
\caption{Controlled \(m_T=2~\mathrm{TeV}\), seed-42 test results for validation-selected 50, 70 and 80 percent signal-efficiency targets, and the fixed 0.99 diagnostic. Yields use Eq.~(\ref{new:eq:expanded}) and the original normalization. $n_B$ is the raw accepted test background count. Thresholds are shown to six decimal places.}\label{new:tab:wp}
\setlength{\tabcolsep}{5pt}
\begin{tabular}{llrrrrrrr}\hline
Model & Target/cut & $s_0$ & $\epsilon_S$(\%) & $\epsilon_B$(\%) & $S$ & $B$ & $n_B$ & $N_{\rm eff,B}$ \\
\hline
XGBoost & 50 & 0.995718 & 49.69 & 0.1490 & 500.9 & 10224.8 & 2 & 1.00 \\
XGBoost & 70 & 0.985641 & 71.46 & 0.2994 & 720.2 & 20536.0 & 6 & 2.02 \\
XGBoost & 80 & 0.960836 & 80.68 & 0.7491 & 813.2 & 51387.2 & 17 & 5.07 \\
XGBoost & 0.99 & 0.990000 & 66.09 & 0.2979 & 666.1 & 20433.6 & 4 & 2.00 \\
GNN & 50 & 0.995847 & 48.76 & 0.0014 & 491.5 & 98.4 & 1 & 1.00 \\
GNN & 70 & 0.977375 & 70.75 & 0.4497 & 713.1 & 30851.2 & 11 & 3.05 \\
GNN & 80 & 0.938002 & 80.65 & 1.1952 & 812.9 & 81989.5 & 21 & 8.07 \\
GNN & 0.99 & 0.990000 & 60.88 & 0.1519 & 613.6 & 10421.7 & 4 & 1.04 \\
\hline\end{tabular}
\end{table}
The accepted background, rather than the integrated AUC, limits the sensitivity interpretation. At the validation-selected 50\% point, the GNN accepts one test background event and no test $t\bar t$ events. Its $B=98.4$ estimate has a counting error estimate of 98.4 events. This is not a validated prediction of a nearly background-free signal region. For the zero accepted $t\bar t$ events among 671 held-out trials, the usual independent-event binomial model gives the one-sided 95\% bound $\epsilon_{t\bar t}<1-0.05^{1/671}=0.004455$. With the fixed original preselection normalization this corresponds to about $3.05\times10^4$ $t\bar t$ events. This is a bound on one component at a fixed cut, not a simultaneous bound on the total background or a substitute for its modeling uncertainties. XGBoost at the same target gives $B=10224.8$ with a counting error estimate of 10204.8 events. At the 0.99 diagnostic, the corresponding estimates are $20433.6\pm14431.7$ and $10421.7\pm10205.7$. These errors describe only the finite stored simulation. \rev{The accepted event counts and sums of squared weights for each process are retained in the revision records and are available from the authors on reasonable request.} \Cref{new:fig:tail} resolves the high-score background tail explicitly; empty bins do not establish a zero physical rate.
\begin{figure}[H]\centering\includegraphics[width=.95\linewidth]{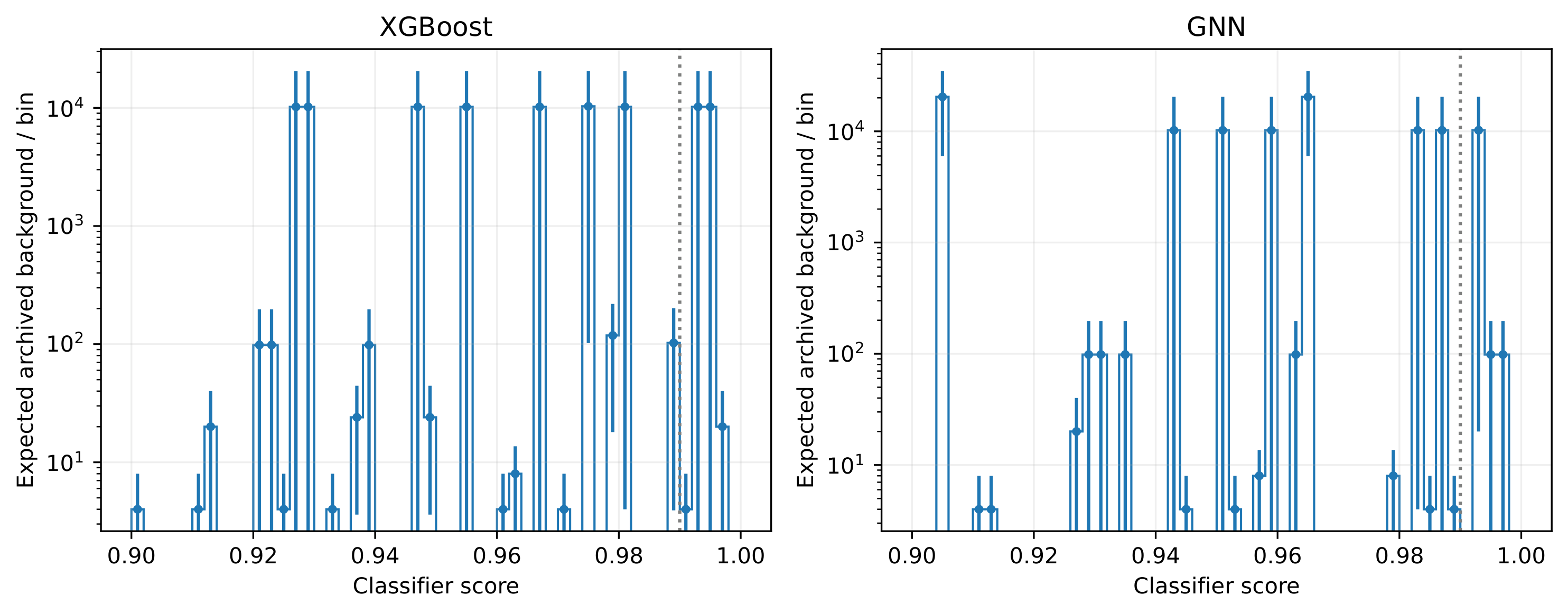}
\caption{Held-out background score tails for the controlled \(m_T=2~\mathrm{TeV}\), seed-42 benchmark under the original normalization. Error bars show square roots of summed squared expanded weights. Empty bins are not evidence of zero background.}\label{new:fig:tail}
\end{figure}
For the independently specified minimum-support diagnostic, validation selects $s_0=0.908685$ for XGBoost and $s_0=0.816639$ for the GNN. Their held-out $N_{\rm eff,B}$ values are 9.15 and 14.18, respectively, and their held-out signal efficiencies are 86.69\% and 87.98\%. In particular, requiring ten effective events in validation does not guarantee ten in test. We retain the frozen thresholds and report this loss of support, rather than adapting them to the test result. These observations, together with the missing background classes, prevent a reliable reach statement from this sample.
\subsubsection{Dependence on signal mass}
The protocol was applied unchanged to all seven available mass samples. \Cref{new:tab:massauc} gives the minimum and maximum test AUC over the three training seeds at each mass. In all 21 paired runs, XGBoost has the larger test AUC; the differences span 0.00024--0.00228. The three-seed ranges are not statistical confidence intervals. The nonmonotonic mass dependence should not be interpreted as a physical acceptance law: each mass uses its own stored signal population and fitted classifier, while the background test population is shared. \rev{The complete working-point results, process yields, training histories, and score arrays for all masses are retained in the revision records and are available from the authors on reasonable request.} No mass-reach contour is derived from these discrimination measurements.
\begin{table}[htbp]\centering\small
\caption{Test AUC ranges across seeds 42, 137 and 2026. Common settings and information inputs are used at every mass.}\label{new:tab:massauc}
\begin{tabular}{rrr}\hline
$m_T$ (TeV) & XGBoost & GNN\\\hline
1.8 & 0.98735--0.98771 & 0.98535--0.98603 \\
2.0 & 0.98562--0.98655 & 0.98428--0.98472 \\
2.2 & 0.99209--0.99231 & 0.99166--0.99173 \\
2.5 & 0.98936--0.98992 & 0.98804--0.98838 \\
2.7 & 0.99069--0.99088 & 0.98998--0.99045 \\
2.9 & 0.99330--0.99337 & 0.99222--0.99263 \\
3.1 & 0.98948--0.98957 & 0.98859--0.98887 \\
\hline\end{tabular}\end{table}

\subsection{Correction of the historical interpretation}\label{new:sec:historical}
Recovery of the original global-GNN script and its numerical summary allows two issues to be identified directly. First, the seven mass samples share identical background records. Their per-mass validation partitions were then pooled with training partitions from other masses. For each mass, 1222 of the 1223 validation background records occur in another mass's training partition. No corresponding test-background overlap with training was found. The issue is therefore validation contamination, affecting early stopping and any validation-based choices; it is not evidence of direct test-background leakage. The new process-stratified splits avoid this construction: each model uses one mass, and the common background records have identical partition assignments across all masses.

Second, the original GNN summary sums the luminosity weights over only the test subset without a sampling correction. At 2 TeV it records $S_{\rm selected}=111.25795$, $B_{\rm selected}=2046.98291$, $S_{\rm test}=201.58618$ and $B_{\rm test}\simeq1348180.25$. The corresponding efficiencies are 55.191\% and 0.15183\%, using denominators from the same test population. \rev{An earlier raster-based reconstruction combined information from the selected test sample with full-sample preselection denominators. The resulting efficiencies are therefore omitted from the revised results and superseded by the event-level summary and controlled working-point results in \cref{new:tab:wp}.}

These unexpanded test sums are not full-luminosity yields. At the nominal 20\% inclusion probability, a factor-five correction illustrates the scale of the problem, but it cannot recover an independently validated tail or missing per-process score records. The maximum recorded value $Z_A=3.835$ consequently does not have the full-luminosity interpretation previously assigned to it. It is not appropriate simply to shift the coupling crossings until they agree with this number.

There is a further algebraic consistency requirement. For fixed positive $B$, the statistical Asimov expression obeys
\begin{equation}
Z_A^2=2[(S+B)\ln(1+S/B)-S],\qquad
\frac{\partial Z_A^2}{\partial S}=2\ln(1+S/B)>0.
\end{equation}
Thus, under one normalization and $S\propto(g^*)^2$, a benchmark below five standard deviations at $g^*=0.4$ cannot reach five at a smaller coupling. The old figure and threshold table cannot both be interpreted under those assumptions. The original images and tables are retained for traceability, but their coupling-contour and mass-reach interpretations are withdrawn. A replacement requires verified production normalization, coupling and width conventions, background coverage, and an appropriate likelihood; none is supplied by an algebraic repair of the old table.

\subsubsection{Correlations of the complete input set}
\rev{The matrices below contain all 45 nonconstant node, edge, and global inputs. Constant role indicators are omitted because their Pearson correlation coefficients are undefined. For compactness, the axes use \(\ell_{p_T}\equiv\log(1+p_T/\mathrm{GeV})\) and \(\ell_m\equiv\log[1+\max(m,0)/\mathrm{GeV}]\). The signal and four-process background correlations are calculated with physical event weights after the stored pre-classifier selection.}
\begin{figure}[!htbp]\centering\includegraphics[width=.95\linewidth]{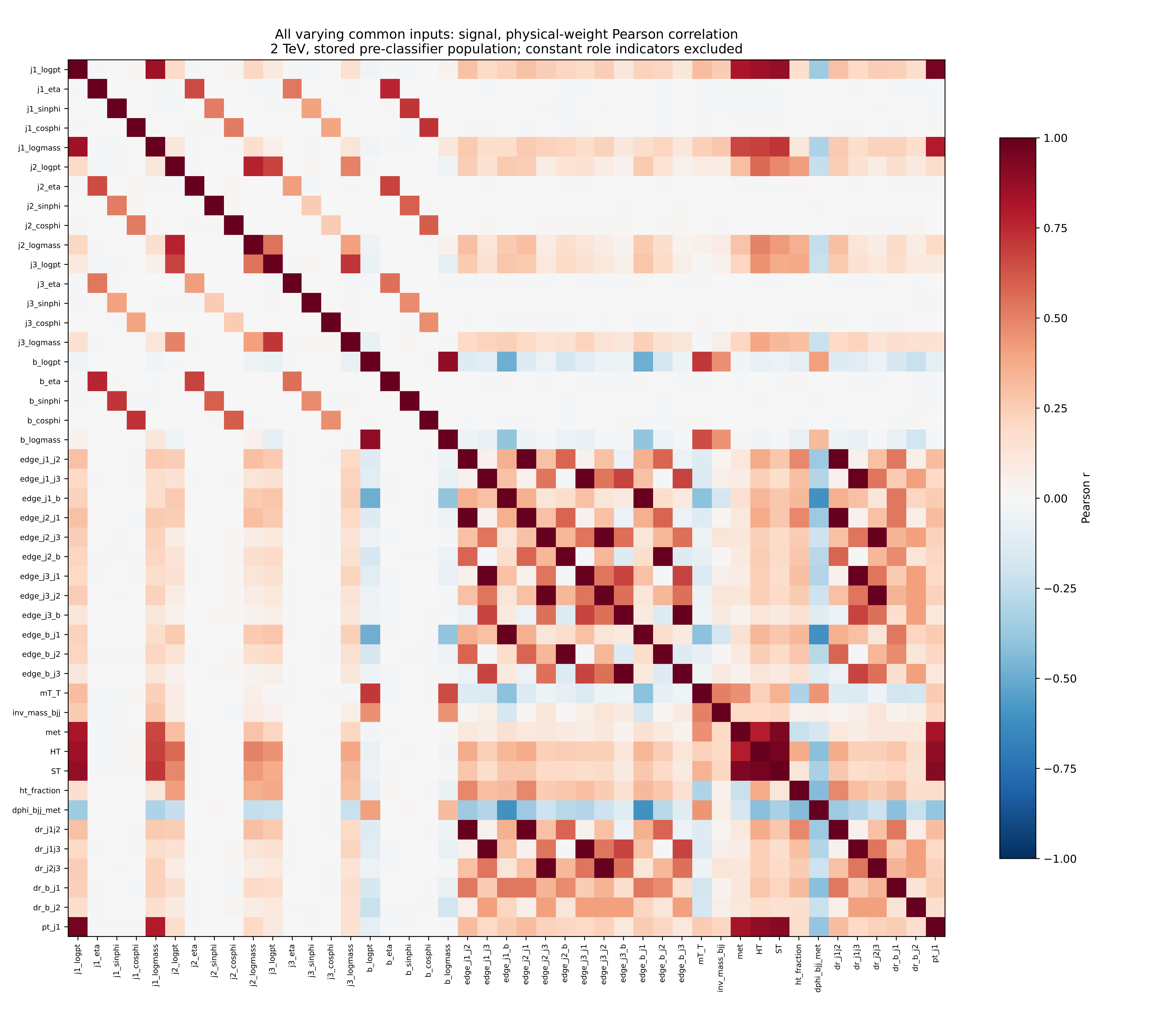}\caption{Complete varying-input correlation matrix for the 2 TeV signal.}\label{new:fig:allcorrsignal}\end{figure}
\begin{figure}[!htbp]\centering\includegraphics[width=.95\linewidth]{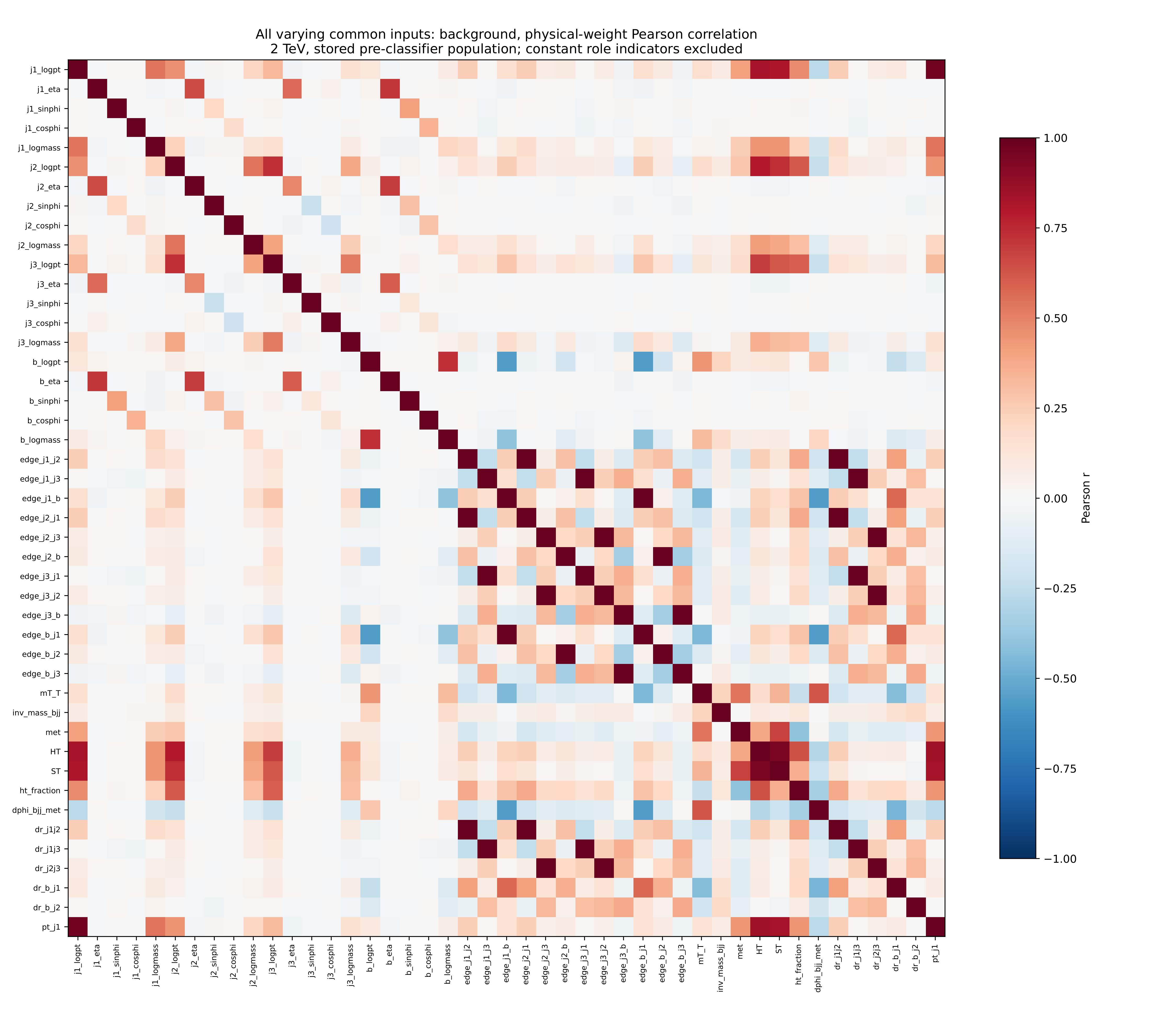}\caption{Complete varying-input correlation matrix for the four-process background mixture.}\label{new:fig:allcorrbackground}\end{figure}

\par}
\FloatBarrier

\section{Quantitative Comparison with Existing Studies}
\label{sec:comparison}

\rev{Because the coupling-rescaling contours and threshold crossings are excluded from the revised interpretation, no numerical mass-reach comparison is derived from them.}

\rev{For context, \Cref{tab:existing_searches} summarizes relevant published searches. Their results cannot be compared numerically with the benchmark sensitivities reported here because the collision energies and luminosities differ, the experimental analyses constrain model-dependent cross sections or \(\kappa_T\) benchmarks, and their likelihoods are constrained by data. A conversion to \(g^*\) would require a consistent UFO implementation, width prescription, branching fractions, and production cross sections.}

\begin{table}[!htbp]
  \centering
  \caption{{Selected published searches for singly produced vector-like \(T\) quarks. Quoted values are observed 95\% confidence-level results and are included as experimental context for the channel studied here; no direct numerical conversion to the present \(g^*\) convention is attempted.}}
  \label{tab:existing_searches}

  \scriptsize
  \newcommand{\resulttablecell}[2]{%
    \parbox[t]{#1}{\raggedright\arraybackslash #2}%
  }
  \renewcommand{\arraystretch}{1.15}

  \begin{tabular}{|l|l|l|l|}
  \hline
  \resulttablecell{2.1cm}{Study} &
  \resulttablecell{2.2cm}{Data set} &
  \resulttablecell{4.4cm}{Relevant feature} &
  \resulttablecell{6.0cm}{Published result} \\
  \hline

  \resulttablecell{2.1cm}{CMS \cite{CMS2022tZMET}} &
  \resulttablecell{2.2cm}{13~TeV, 137~fb\(^{-1}\)} &
  \resulttablecell{4.4cm}{Jets plus missing momentum, \(T\to tZ\)} &
  \resulttablecell{6.0cm}{Narrow-width \(\sigma\mathcal{B}\) limits of 602--15~fb for \(m_T=0.6\)--1.8~TeV; a 5\%-width singlet benchmark is excluded below 0.98~TeV.} \\
  \hline

  \resulttablecell{2.1cm}{ATLAS \cite{ATLAS2023tZtH}} &
  \resulttablecell{2.2cm}{13~TeV, 139~fb\(^{-1}\)} &
  \resulttablecell{4.4cm}{Combined \(Ht\) and \(Zt\) final states} &
  \resulttablecell{6.0cm}{Singlet couplings \(\kappa_T>0.53\) are excluded for all tested masses below 2.3~TeV; \(\kappa_T=0.35\) is excluded at 1.6~TeV.} \\
  \hline

  \resulttablecell{2.1cm}{ATLAS \cite{ATLAS2024MonoTopXGB}} &
  \resulttablecell{2.2cm}{13~TeV, 139~fb\(^{-1}\)} &
  \resulttablecell{4.4cm}{Boosted top plus missing momentum; XGBoost top tagger} &
  \resulttablecell{6.0cm}{\(m_T<1.8~\mathrm{TeV}\) excluded for \(\kappa_T=0.5\) and \(\mathcal{B}(T\to tZ)=25\%\).} \\
  \hline

  \resulttablecell{2.1cm}{CMS \cite{CMS2024tHtZ}} &
  \resulttablecell{2.2cm}{13~TeV, 138~fb\(^{-1}\)} &
  \resulttablecell{4.4cm}{All-hadronic \(tH/tZ\) final state} &
  \resulttablecell{6.0cm}{Cross-section limits of 1260--68~fb over \(m_T=0.6\)--1.2~TeV.} \\
  \hline

  \resulttablecell{2.1cm}{ATLAS \cite{ATLAS2025SingleTCombination}} &
  \resulttablecell{2.2cm}{13~TeV, 139~fb\(^{-1}\)} &
  \resulttablecell{4.4cm}{Combination of single-\(T\) \(Ht\) and \(Zt\) searches} &
  \resulttablecell{6.0cm}{Singlet benchmark with \(\kappa_T=0.5\) excluded below 2.1~TeV.} \\
  \hline
  \end{tabular}
  \end{table}

%%%%%%%%%%%%%%%%%%%%%%%%%%%%%%%%%%%%%%%%%%%%%%%%%%%%%%%%%%%%%%%%%%%%%%%%
\section{Conclusion}
\label{sec:conclusion}
\rev{The controlled comparison in \cref{new:sec:analysis} uses identical physical information, fixed process-stratified partitions, seven mass points, and three training seeds. XGBoost yields the larger held-out AUC in all 21 paired runs, with differences of \(2.4\times10^{-4}\)--\(2.28\times10^{-3}\). This result applies only to the specified configurations and does not establish an architecture-wide optimum. The normalization audit supersedes the earlier raster-derived operating points, inferred post-classifier efficiencies, and associated sensitivity interpretation. The small number of accepted background events and the omission of relevant background processes preclude a validated physical reach estimate.}

%%%%%%%%%%%%%%%%%%%%%%%%%%%%%%%%%%%%%%%%%%%%%%%%%%%%%%%%%%%%%%%%%%%%%%%%
We have audited a feasibility study of single production of a singlet top partner in the hadronic \(tZ\) channel at the 14~TeV HL-LHC. \rev{The independently configured historical pipelines yield test-AUC ranges of 0.976--0.983 for XGBoost and 0.983--0.990 for the GNN. Because they use different feature representations, these values are reported only as pipeline-specific results. The controlled common-input study provides a matched comparison of the specified configurations; its small numerical differences do not establish a general ranking of the classifiers.}

The statistical Asimov significance recorded with the original normalization spans 0.174--3.355 for XGBoost and 0.234--3.835 for the GNN. \rev{A 20\% normalization uncertainty on the summed four-process background reduces the corresponding maxima to 0.258 and 0.416. These values quantify the conditional sensitivity of the simulated background model and are not interpreted as experimental exclusion or discovery limits.} \rev{The internally inconsistent coupling-rescaling contours and threshold table are excluded from the revised interpretation. The quantitative sensitivity discussion is therefore restricted to the directly simulated benchmark at \(g^*=0.4\) and \(R_L=0.5\).}

\rev{Within this scope, the available samples provide a benchmark for classifier performance and conditional signal sensitivity at \(g^*=0.4\) and \(R_L=0.5\). An experimental HL-LHC limit would require the missing background components and a likelihood treatment of normalization and shape uncertainties, ideally constrained by control regions. We therefore report results only for the simulated background model and do not extrapolate the recorded significances to experimental exclusion or discovery limits.}
%%%%%%%%%%%%%%%%%%%%%%%%%%%%%%%%%%%%%%%%%%%%%%%%%%%%%%%%%%%%%%%%
\section{Statements and Declarations}
\label{sec:declarations}
\subsection{Acknowledgments}
The authors thank Mr. Kamran Ahmad (AM Physics, National Centre for Physics, Pakistan) for valuable discussions and time.
\subsection{Funding}
This research did not receive any grant from funding agencies in the public, commercial, or not-for-profit sectors.
\subsection{Competing Interests}
The authors have no relevant financial or non-financial interests to disclose.
\subsection{Data and code availability}
\rev{The processed artifacts underlying the controlled machine-learning comparison---including preprocessing constants, split identifiers, model files, training histories, event-level scores, and numerical working-point results---are available from the authors on reasonable request. The original generator cards, Delphes card, and complete upstream Monte Carlo generation and reconstruction chain are unavailable.} 

%%%%%%%%%%%%%%%%%%%%%%%%%%%%%%%%%%%%%%%%%%%%%%%%%%%%%%%%%%%%%%%%%%%%%%%%
% CPC-2026-0641 normalization revision: original threshold appendix retained in source but excluded from compiled manuscript.
\iffalse
\appendix
\section{Original Statistical Threshold Table}
\label{sec:appendix}
\begin{table}[H]
\centering
\small
\caption{Original $g^*$ crossings of the statistical-only thresholds $Z_A=2$ and $Z_A=5$ at each $m_T$. Entries marked ``--'' indicate that the crossing is not reached within the scanned $g^*$ range. The $R_L=0$ reweighting is not reproducible from the surviving files, and these entries must not be interpreted as confidence limits.}
\label{tab:gstar_thresholds}
\begin{tabular}{|c|c|c|c|c|c|}
\hline
$R_L$ & $m_T$ [GeV] & XGB $g^*_{Z_A=2}$ & XGB $g^*_{Z_A=5}$ & GNN $g^*_{Z_A=2}$ & GNN $g^*_{Z_A=5}$ \\
\hline
0.0 & 1800 & 0.17 & 0.27 & 0.16 & 0.26 \\
0.0 & 2000 & 0.22 & 0.35 & 0.20 & 0.32 \\
0.0 & 2200 & 0.28 & 0.44 & 0.26 & 0.40 \\
0.0 & 2500 & 0.39 & -- & 0.35 & -- \\
0.0 & 2700 & 0.49 & -- & 0.43 & -- \\
0.0 & 2900 & -- & -- & -- & -- \\
0.0 & 3100 & -- & -- & -- & -- \\
0.5 & 1800 & 0.21 & 0.33 & 0.20 & 0.31 \\
0.5 & 2000 & 0.27 & 0.43 & 0.25 & 0.39 \\
0.5 & 2200 & 0.34 & -- & 0.31 & 0.49 \\
0.5 & 2500 & 0.48 & -- & 0.43 & -- \\
0.5 & 2700 & -- & -- & -- & -- \\
0.5 & 2900 & -- & -- & -- & -- \\
0.5 & 3100 & -- & -- & -- & -- \\
\hline
\end{tabular}
\end{table}
\fi

%\bibliographystyle{plainnat}
%\bibliography{references}

%%%%%%%%%%%%%%%%%%%%%%%%%%%%%%%%%%%%%%%%%%%%%%%%%%%

\end{document}